\documentclass[a4paper,11pt]{article}
\usepackage{jcappub} % for details on the use of the package, please see the JINST-author-manual
\usepackage{natbib}
\usepackage[T1]{fontenc}
\usepackage{xcolor}
\usepackage{booktabs}
\usepackage{graphicx}
\usepackage{subcaption}
\title{Self-Similar Mass Spectra of Hierarchical Black Hole Mergers}

\author[a]{Liam Blum, }
\author[a,b]{Stefano Profumo, }
\affiliation[a]{Department of Physics, University of California, Santa Cruz \\Santa Cruz, CA, 95064, USA}
\affiliation[b]{Santa Cruz Institute for Particle Physics,\\Santa Cruz, CA, 95064, USA}
\emailAdd{profumo@ucsc.edu}
\author[a]{Ryan Schantz, }
\author[a]{Quoc Ha Tran}

\hypersetup{
  pdftitle={Self-Similar Mass Spectra of Hierarchical Black Hole Mergers},
  pdfauthor={Liam Blum, Stefano Profumo, Ryan Schantz, Quoc Ha Tran}
}

\begin{document}

\newcommand{\msun}{M_{\odot}}
\newcommand{\kms}{\text{km}~\text{s}^{-1}}
\newcommand{\spr}[1]{{\color{red}\bf[SP:  {#1}]}}

\abstract{Black holes that merge repeatedly carry a record of that history in their mass
distribution. We treat hierarchical merging as a coagulation problem, in which a single kernel
encodes how the merger rate depends on the masses involved and on cosmic time, and we derive the
late-time spectra that such populations approach. The framework yields closed scaling laws, an
exactly solvable benchmark, and a classification of merger environments by a single exponent
measuring how strongly mergers favor or suppress massive participants. Applying it to primordial
black holes requires care in translating published merger rates into kernel form; done
correctly, three of the four standard binary-formation channels map onto superlinear kernels and
are therefore candidates for runaway growth, in which the heaviest objects would dominate and no
steady mass-conserving spectrum exists; establishing physical gelation, however, requires the
full population-dependent kernel and its finite coagulation history, neither of which we settle
here. Only the early three-body channel maps into the nongelling regime and admits slow
self-similar growth. Numerical solutions across thirteen suppressive, nongelling kernels show
that the shape of the spectrum is not fixed by the scaling exponent alone,
and that accounting for the energy radiated at each merger measurably changes the high-mass
cutoff and drains the population's total black-hole mass.}

\maketitle
\flushbottom

%=====================================================================
\section{Introduction}\label{sec:intro}
%=====================================================================

The hierarchical merger of black holes (BHs) is a cornerstone of modern gravitational-wave
astronomy. Since the first detection of gravitational waves from a binary black hole (BBH) merger
by the Advanced LIGO detectors in 2015~\cite{Abbott:2016blz}, the LIGO-Virgo-KAGRA (LVK)
collaboration has cataloged hundreds of compact binary coalescences through successive observing
runs~\cite{LIGOScientific:2018mvr, Abbott:2020niy, LIGOScientific:2021djp}, with
catalogs continuing to expand the observed
population~\cite{Nitz:2021zwj, Olsen:2022pin, Wadekar:2023gea, Abac:2025saz}.
These observations have revealed a rich landscape: a power-law primary mass spectrum with
structure near $\sim 35\,M_\odot$, a paucity of objects in the pair-instability
gap~\cite{Fishbach:2017zga, Talbot:2018cva, Abbott:2021iab}, and mounting evidence for
a subpopulation of black holes whose masses challenge standard stellar-evolution
theory~\cite{Abbott:2020tfl, Abbott:2021iab}. Understanding how repeated BH mergers shape
these observed distributions is a central open problem.

A pair-instability supernova (PISN) mechanism is expected to suppress the formation of
stellar-remnant BHs in the mass range $\sim 50$--$120\,\msun$~\cite{Woosley:2016hmi,
Farmer:2019jed, Marchant:2018kun}. Yet gravitational-wave events such as GW190521---whose
primary component mass lies well within the PISN gap~\cite{Abbott:2020mjq,
Abbott:2020tfl}---demand alternative formation pathways. Hierarchical mergers, in which
second- and higher-generation BHs are assembled from earlier coalescences, naturally populate
the gap~\cite{Gerosa:2017kvu, Fishbach:2017zga, Mapelli:2021syv}, provided the host
environment retains merger products against gravitational-wave recoil
kicks~\cite{Lousto:2008dn, Campanelli:2007ew}. For a comprehensive review of hierarchical
BH mergers and their gravitational-wave signatures, see~\cite{Gerosa:2021hsc}. In the context
of primordial black holes (PBHs), hierarchical mergers additionally modify the initial mass
function set at formation, introducing a time-dependent component to what is otherwise a
purely cosmological relic distribution~\cite{Carr:2020gox, Carr:2016drx, Sasaki:2018dmp}.

Despite the observational maturity of the field, the theoretical description of BH mass
spectra under iterated mergers remains largely numerical. Binary-population codes such as
\textsc{COMPAS}~\cite{Stevenson:2017tfq, Riley:2022kid} and
\textsc{COSMIC}~\cite{Breivik:2019lmt}, together with cluster-dynamics tools such as
\textsc{CMC}~\cite{Rodriguez:2022tso}, model the relevant stellar evolution and dynamical
processes in detail. Cluster calculations show that hierarchical growth depends sensitively on
remnant retention and therefore on the host escape velocity~\cite{Mapelli:2021syv,
Zevin:2022bbh, Rodriguez:2019huv}. Semi-analytic models of hierarchical mass
buildup~\cite{Baibhav:2020xdf, Antonini:2018auk} have elucidated the conditions under which
runaway growth and intermediate-mass BH (IMBH) formation are possible~\cite{Arca-Sedda:2021fkf}.
AGN-disk calculations provide a qualitatively distinct channel in which gas torques, migration,
and repeated encounters reshape the merging population~\cite{Yang:2019cbr,
Bellovary:2016migration, McKernan:2014agn}. At the same time, population-level inference
applied to LVK data~\cite{Abbott:2021iab, Edelman:2021fik} has revealed mass-spectrum
substructure whose physical origin is not yet understood.

What remains comparatively underdeveloped is a systematic \emph{analytic} treatment of
how the BH mass distribution evolves under hierarchical mergers, i.e., one that derives
explicit scaling laws and self-similar solutions rather than relying on simulation. This gap
is not merely academic: analytic solutions expose the universality classes of merger dynamics,
establish the correct mapping between kernel parameters and observable features of the mass
spectrum (power-law slopes, high-mass tails, and qualitatively, mass-gap phenomenology), and
provide physically
transparent benchmarks for numerical codes. A complementary analytic framework is particularly
timely given the increasingly precise BH population measurements from current LVK
catalogs~\cite{LIGOScientific:2021djp, Abac:2025saz} and the prospective reach of
next-generation detectors such as LISA~\cite{Colpi:2024lisa}, the Einstein
Telescope~\cite{Punturo:2010zz}, and Cosmic Explorer~\cite{Reitze:2019iox}.

Smoluchowski's coagulation equation~\cite{Smoluchowski:1916, Drake:1972, Aldous:1999},
originally formulated to describe Brownian aggregation of colloidal particles, provides
precisely such a framework. In this mean-field description, BHs of mass $m$ and $m'$ merge
at a rate proportional to a kernel $K(m, m', t)$, and the resulting integro-differential
equation governs the time evolution of the mass number density $c(m,t)$. The coagulation
approach has been applied fruitfully in cosmological contexts: to dark matter halo growth and
the Press--Schechter formalism~\cite{Lacey:1993iv, Benson:2005ms}, to planetesimal
accretion~\cite{Wetherill:1990}, and to dust coagulation in protoplanetary
disks~\cite{Dullemond:2005}. Smoluchowski's equation has also been applied directly to
idealized runaway black-hole merging in dense stellar systems~\cite{Mouri:2002}. The goal
here is different: to develop a common similarity framework for hierarchical astrophysical and
primordial-black-hole merger kernels, while making explicit which conclusions follow from
homogeneity and which require the full kernel.

The mathematical theory of self-similar solutions to the coagulation equation is
well-developed for homogeneous kernels of degree $\lambda$~\cite{Leyvraz:2003, Menon:2004,
Fournier:2005, vanDongen:1988}. Existence of self-similar profiles for a broad class of
homogeneous kernels is established in Ref.~\cite{Fournier:2005}, and convergence toward
self-similarity for solvable kernels in Refs.~\cite{Menon:2004, Menon:2004b}; see also
Ref.~\cite{Aldous:1999}. Homogeneity is the central organizing quantity: kernels with degree
below unity commonly admit mass-conserving scaling states, the marginal degree $\lambda=1$
requires kernel-specific analysis, and sufficiently superlinear kernels can exhibit gelation or
runaway growth~\cite{Leyvraz:2003, Escobedo:2002}. Endpoint and tail exponents, however, are
properties of the full kernel and are not fixed by homogeneity alone.

For primordial black holes specifically, the merger rate is determined by four distinct binary
formation channels: early two-body (E2) and three-body (E3) configurations, which form during
the radiation-dominated era through gravitational decoupling from the Hubble flow, and late
two-body (L2) and three-body (L3) configurations, which form through gravitational-wave capture
and multi-body encounters in dark matter
halos~\cite{Nakamura:1997sm, Sasaki:2016jop, Bird:2016dcv, Ali-Haimoud:2017rtz,
Raidal:2018bbj, Raidal:2024bmm}. Each channel yields a distinct mass and time dependence in
the merger rate, which we translate into the kernel parameters $(\alpha,\beta,\delta)$ of our
formalism.

A feature that distinguishes black-hole coagulation from its classical counterparts deserves
emphasis at the outset. In colloidal aggregation, planetesimal accretion, or dust growth, mass
is conserved exactly: two units combine into one of their summed mass. Black-hole mergers are
not of this type. A comparable-mass binary radiates several percent of its total mass in
gravitational waves, so the remnant is lighter than the sum of its progenitors, and the total
black-hole mass of the population decays monotonically as merging
proceeds~\cite{Barausse:2012mass}. This is not a small correction to be absorbed into a
redefinition: mass conservation is the assumption underpinning the standard self-similar
theory, and it is what fixes the amplitude exponent through $\theta=2z$. Once it fails, the
characteristic scale and the amplitude cease to be locked together and must be determined
separately from the non-conservative evolution. Because the radiated fraction is itself
mass-ratio and spin dependent, the effect does not simply rescale the spectrum; it reshapes it.
We therefore treat the conservative problem as an analytic baseline and the radiating problem as
a distinct dynamical system, and we quantify the difference by solving both across the same
family of kernels. The comparison turns out to be systematic: at every kernel we sample,
radiative losses raise the fitted cutoff exponent, widen the characteristic scale, and shift the
low-mass profile, while the total black-hole mass drains at a rate set by the kernel itself.

In this paper we apply Smoluchowski's coagulation equation with the kernel class
\begin{equation}
K(m,m',t) = (m+m')^{-\alpha}(mm')^{-\beta}\,t^{-\delta}, \qquad \alpha,\beta,\delta\in\mathbb{R}.
\label{eq:master_kernel_intro}
\end{equation}
The parameters $\alpha$, $\beta$, and $\delta$ are phenomenological exponents that encode,
respectively, dependence on total mass, component-mass bias, and explicit time evolution. The
coagulation literature often studies homogeneous product or sum kernels as analytically
tractable representatives~\cite{vanDongen:1988, Leyvraz:2003}. In black-hole applications,
gravity-driven cross-sections, mass segregation, dynamical friction, binary hardening, and
three-body interactions motivate nontrivial mass dependence~\cite{Spitzer:1987,
Rodriguez:2019huv, Antonini:2018auk}, but they do not uniquely select the two-power
parametrization in Eq.~\eqref{eq:master_kernel_intro}. We therefore use
Eq.~\eqref{eq:master_kernel_intro} as a controlled homogeneous model class and map specific
rate formulae onto it only when their mass dependence is explicit.

Our main results are:
\begin{enumerate}
    \item The mass-conserving similarity form $c(m,t)=s(t)^{-2}\Phi(m/s(t))$, with
    $s\propto T^{1/(1-\lambda)}$ in the cumulative coagulation time $T(t)=\int^t t'^{-\delta}dt'$,
    reducing for $\delta<1$ to $s\propto t^z$ with $z=(1-\delta)/(1+\alpha+2\beta)$; together
    with the closed nonlinear similarity equation for $\Phi$ and an exact constant-kernel
    benchmark.
    \item The conversion between the $M$--$\eta$ parametrization of the PBH literature and the
    coagulation kernel. Because quoted merger rates are rate \emph{densities} rather than
    per-pair coefficients, the conversion carries an extra factor $m_1m_2=\eta M^2$; omitting
    it shifts the homogeneity degree by two units.
    \item The resulting classification of the PBH pathways: E2, L2 and L3 map onto superlinear
    kernels ($\lambda>1$) and are candidates for runaway growth, while E3 alone maps into the
    nongelling, mass-conserving similarity regime, with $z\simeq0.0305$--$0.0716$.
    \item Numerical similarity profiles across thirteen kernels, showing that equal-homogeneity
    kernels can have measurably different shapes: total-mass suppression and product-mass
    suppression are not interchangeable at fixed $\lambda$.
    \item A systematic quantification of gravitational-wave mass loss. Repeating every
    calculation with a retained fraction $L=0.95$, we find that the cutoff exponent $q$ rises at
    \emph{every} kernel sampled, by about $0.11$ along the total-mass axis and by up to $0.14$
    along the product-mass axis, and that the characteristic scale $\xi_0$ rises at every kernel
    as well, while the response of the algebraic exponent $p$ changes sign across the sampled
    range. Because the first moment now decays, the conservative relation $\theta=2z$ no longer
    holds and the amplitude exponent must be measured rather than inferred.
\end{enumerate}

The paper is organized as follows. Section~\ref{sec:smoluch} introduces the coagulation equation,
the homogeneous kernel class, and gravitational-wave mass loss.
Sections~\ref{sec:astrobh} and~\ref{sec:pbh} map representative astrophysical and PBH merger
channels onto the kernel notation. Section~\ref{sec:scaling} derives the similarity structure,
contains the exact constant-kernel benchmark, and states the dynamical classification.
Section~\ref{sec:numerical} describes the numerical method and compares the elastic and
mass-loss profiles. Section~\ref{sec:conclusions} summarizes the conclusions and open problems.

%=====================================================================
\section{Smoluchowski's Coagulation Equation}\label{sec:smoluch}
%=====================================================================

\subsection{Master equation and kernel class}

We consider a population of compact objects with mass $m\ge m_0$, where $m_0$ is a
physical lower bound set by stellar-evolution considerations (e.g.\ the Chandrasekhar or
TOV mass scale) for astrophysical BHs, or by the hypothesized PBH mass spectrum cutoff.
Let $c(m,t)$ denote the differential number density per unit mass, so that
\begin{equation}
N(t) = \int_{m_0}^{\infty} c(m,t)\,dm,
\qquad
M_1(t) = \int_{m_0}^{\infty} m\,c(m,t)\,dm
\label{eq:N_M1_def}
\end{equation}
are the total number density and total mass density, respectively.  In the absence of
external sources or sinks, and neglecting energy radiated during mergers, the idealized
binary-merger process conserves total black-hole mass, $dM_1/dt=0$.
Section~\ref{sec:mass_loss} relaxes this approximation to include gravitational-wave emission.

We describe mergers in a mean-field fashion by the Smoluchowski coagulation
equation~\cite{Smoluchowski:1916,Drake:1972,Aldous:1999}
\begin{align}
\partial_t c(m,t)
&=
\frac{1}{2}
\int_{m_0}^{m-m_0}
K(m-m',m',t)\,
c(m-m',t)\,c(m',t)\,dm'
\nonumber\\
&\quad
- c(m,t)
\int_{m_0}^{\infty}
K(m,m',t)\,c(m',t)\,dm',
\label{eq:smoluchowski}
\end{align}
where the first (gain) term accounts for mergers of masses $m'$ and $m-m'$ producing a
remnant of mass $m$, while the second (loss) term accounts for the depletion of mass-$m$
objects through mergers with any companion. Equivalently, the merger rate per unit volume
with progenitor masses in $dm_1\,dm_2$ is
\begin{equation}
\frac{dR}{dm_1\,dm_2}=\frac{1}{2}\,K(m_1,m_2,t)\,c(m_1,t)\,c(m_2,t),
\label{eq:rate_kernel_relation}
\end{equation}
a relation we will need in Sec.~\ref{sec:pbh_conversion} to convert published merger-rate
formulae into kernels.

In what follows we work with the kernel class
\begin{equation}
K(m,m',t)
=
(m+m')^{-\alpha}\,(m m')^{-\beta}\,t^{-\delta},
\qquad \alpha,\beta,\delta\in\mathbb{R},
\label{eq:general_kernel}
\end{equation}
where the explicit factor $t^{-\delta}$ allows for cosmological dilution or environmental
evolution.  Under the rescaling $(m,m')\to(am,am')$, the kernel transforms as
$K(am,am',t) = a^{-(\alpha+2\beta)}K(m,m',t)$, so its degree of homogeneity in mass is
\begin{equation}
\lambda \equiv -(\alpha+2\beta).
\label{eq:homogeneity}
\end{equation}

%---------------------------------------------------------------------
\subsection{Gravitational-wave mass loss}
\label{sec:mass_loss}
%---------------------------------------------------------------------

A binary merger of masses $m_1$ and $m_2$ radiates a fraction of the total energy in
gravitational waves. We write the remnant mass as
\begin{equation}
M_{\rm final}=L(m_1+m_2),\qquad L\equiv1-\epsilon_{\rm GW},
\label{eq:mass_loss_factor}
\end{equation}
where $L$ depends on mass ratio and spin. This prescription applies to mergers of both
stellar-origin and primordial black holes, and describes mass loss \emph{at} a merger rather
than continuous mass evolution between mergers; individual masses are held fixed between
coalescences, and Hawking evaporation and accretion are outside the scope of this study.
Numerical-relativity remnant fits show that a few percent of the total mass is typically
radiated for comparable-mass BBH mergers, so $L\simeq0.95$ is a useful representative value
rather than a universal constant~\cite{Barausse:2012mass}.

For constant $L$, the gain term can be written exactly as
\begin{align}
\partial_t c(m,t)
&=\frac{1}{2L^2}\int_{Lm_0}^{m-Lm_0}
K\!\left(\frac{m-m'}{L},\frac{m'}{L},t\right)
 c\!\left(\frac{m-m'}{L},t\right)c\!\left(\frac{m'}{L},t\right)\,dm'
\nonumber\\
&\quad-c(m,t)\int_{m_0}^{\infty}K(m,m',t)c(m',t)\,dm'.
\label{eq:smoluchowski_mass_loss}
\end{align}
The loss term is unchanged because it counts removal of a progenitor of mass $m$; the gain
term is shifted because the remnant retains only a fraction $L$ of the progenitor mass sum.

The structural consequence is that the first moment is no longer conserved. Multiplying by
$m$ and integrating gives, for a closed system,
\begin{equation}
\frac{dM_1}{dt}
=-\frac{1-L}{2}\int_{m_0}^{\infty}\!dm_1\int_{m_0}^{\infty}\!dm_2\,
(m_1+m_2)K(m_1,m_2,t)c(m_1,t)c(m_2,t).
\label{eq:gw_mass_moment_loss}
\end{equation}
Consequently the mass-conserving relation $\theta=2z$ derived in Sec.~\ref{sec:scaling} for
$L=1$ does not carry over to $L<1$, even though the mass homogeneity of the kernel is
unchanged. We therefore treat the conservative $L=1$ problem as the analytic baseline and
$L<1$ as a distinct non-conservative problem in the numerical analysis: profile shapes may
remain close for modest mass loss, but the amplitude and characteristic-scale exponents must
be measured rather than imported from the conservative solution.

\paragraph{Effect on the support.}
A merger of two objects with $m_1,m_2\ge m_0$ produces a remnant with $M_{\rm final}\ge2Lm_0$,
so for $L\ge1/2$ mergers do not create objects below the original cutoff $m_0$. The population
does not, however, acquire a new global lower cutoff at $2Lm_0$: unmerged first-generation
objects remain at $m\simeq m_0$, and higher-generation bounds depend on the merger tree (for
equal-generation pairings, $(2L)^nm_0$). A single generation-independent shifted support should
therefore not be imposed on the full population.

%---------------------------------------------------------------------
\subsection{Qualitative behavior of $K(m,m')$}
\label{sec:kernel_qualitative}
%---------------------------------------------------------------------

The functional form of $K$ determines which mass scale dominates the merger hierarchy.
For the general class~\eqref{eq:general_kernel}, the ratio $K(m,m')/K(m_0,m_0)$ at fixed
$m'>m_0$ decreases with $m$ when $\alpha+\beta\geq0$, so that larger-mass mergers are suppressed
relative to small-mass mergers; outside this quadrant the sign of the trend depends on the mass
ratio. This is the physically relevant regime for the isolated-binary populations discussed in
Sec.~\ref{sec:astrobh}, whose per-pair merger probability falls with total mass, as opposed to
the mass-favoring regime characteristic of dynamically assembled clusters, where dynamical
friction instead concentrates and enhances interactions among the most massive objects. Four
qualitative behaviors arise:
\begin{itemize}
\item \textbf{Constant kernel} ($\alpha=\beta=0$, $K=1$, $\lambda=0$): uniform merging rate;
      the exact monodisperse solution is geometric in cluster number and approaches an
      exponential self-similar profile at late times
      (Sec.~\ref{subsec:constant-kernel-exact}).
\item \textbf{Additive kernel} ($K\propto m+m'$, i.e.\ $\alpha=-1$, $\beta=0$, $\lambda=1$):
      a marginal, exactly solvable kernel whose scaling behavior is
      kernel-specific~\cite{Leyvraz:2003, Menon:2004b}.
\item \textbf{Multiplicative kernel} ($K\propto mm'$, i.e.\ $\alpha=0$, $\beta=-1$,
      $\lambda=2$): a classical gelling kernel in which a macroscopic component forms in
      finite coagulation time~\cite{Leyvraz:2003, Escobedo:2002}.
\item \textbf{Suppressive kernels} ($\alpha,\beta>0$, $\lambda<0$): high-mass encounters are
      progressively suppressed. The numerical solutions of Sec.~\ref{sec:numerical} exhibit
      exponential or stretched-exponential high-mass cutoffs, with a cutoff exponent that is an
      empirical property of the full kernel.
\end{itemize}

%---------------------------------------------------------------------
\subsection{Low-mass cutoff and support properties}
\label{sec:cutoff_support}
%---------------------------------------------------------------------

We impose the physical cutoff $c(m,t)=0$ for $m<m_0$.  Because
Eq.~\eqref{eq:smoluchowski} involves only the \emph{addition} of masses and no
fragmentation, the support of $c(\cdot,t)$ is non-decreasing: if $c(m,t_*)=0$ for all
$m<m_0$ at some time $t_*$, then the gain term vanishes identically there (since both
arguments $m'$ and $m-m'$ would need to be $\geq m_0$, but their sum would then be
$\geq 2m_0 > m$), and the loss term also vanishes.  Hence $\partial_t c(m,t_*)=0$ for
$m<m_0$, and no objects below $m_0$ are ever dynamically generated, so that
$\mathrm{supp}\,c(\cdot,t)\subseteq[m_0,\infty)$ for all $t\ge0$. This cutoff remains
important in the similarity limit because it sets the finite-time lower boundary
$\xi_{\min}=m_0/s(t)$, but it does not by itself fix a unique endpoint power law for the
rescaled profile.

%=====================================================================
\section{Astrophysical Black Hole Merger Kernels}\label{sec:astrobh}
%=====================================================================

The merger rate of astrophysical binary black holes depends on the component masses
$m_1, m_2$ and on time, capturing the probability of coalescence over cosmic time as a
function of environment, hierarchical buildup, and intrinsic BH properties.  In this section
we map representative astrophysical environments onto the kernel
class~\eqref{eq:general_kernel} and collect the resulting $(\alpha,\beta,\delta)$ values.

%---------------------------------------------------------------------
\subsection{Kernel forms in astrophysical environments}
%---------------------------------------------------------------------

For astrophysical scenarios we use the one-parameter specialization
\begin{equation}
K(m,m') = K_0\,(m+m')^{-\alpha},
\label{eq:astro_kernel}
\end{equation}
as an illustrative effective kernel, with the normalization $K_0$ set by the environment.
Dense stellar systems provide qualitative motivation for nontrivial total-mass dependence
through gravitational focusing, mass segregation, dynamical friction, and binary
hardening~\cite{Spitzer:1987, Rodriguez:2019huv}, but these processes do not imply a
universal value of $\alpha$.

For isolated binaries, even the gravitational-wave inspiral factor is not a universal kernel.
For a circular binary with fixed initial semi-major axis $a_0$, Peters' result gives
\begin{equation}
t_{\rm GW}(a_0)=\frac{5c^5 a_0^4}{256G^3m_1m_2(m_1+m_2)},
\label{eq:peters_circular_time}
\end{equation}
so at fixed $a_0$ the inverse inspiral time scales as $m_1m_2(m_1+m_2)$~\cite{Peters:1964}.
After marginalizing over the initial separation, eccentricity, stellar progenitor distribution,
mass transfer, common-envelope evolution, and selection effects, however, no unique power of
$m_1$ and $m_2$ follows~\cite{Dominik:2012kk}. We therefore do not identify the chirp mass
itself with a merger kernel; any isolated-binary kernel used in Eq.~\eqref{eq:general_kernel}
must be understood as a phenomenological approximation to the full population model.

%---------------------------------------------------------------------
\subsection{Environment-specific regimes}
%---------------------------------------------------------------------

\paragraph{Dense stellar clusters.}
In globular and nuclear star clusters, retained merger remnants can participate in repeated
coalescences and build up more massive black holes~\cite{Antonini:2018auk,
Rodriguez:2019huv, Mapelli:2021syv}. Nuclear star clusters, whose escape speeds exceed those
of globular clusters, retain a larger fraction of merger products and are correspondingly more
favorable to hierarchical growth~\cite{Mapelli:2021syv, Zevin:2022bbh}; we represent this
mass-favored merging by a negative $\alpha$,
\begin{equation}
K(m_1,m_2) \approx 10^{-10}
\left(\frac{m_1+m_2}{10\,\msun}\right)^{0.5}\!{\rm Myr}^{-1},
\label{eq:cluster_kernel}
\end{equation}
i.e.\ $\alpha=-0.5$, $\beta=0$, hence $\lambda=+0.5$ (subcritical). The normalization
$10^{-10}\,{\rm Myr}^{-1}$ is illustrative only: a coagulation kernel multiplying two number
densities must carry inverse volume-time units, and no attempt is made here to calibrate an
absolute rate; only the mass-dependence exponent $\alpha$ enters the classification of
Table~\ref{tab:astro_kernels}. Globular clusters, with
$v_{\rm esc}\sim50\,\kms$ and correspondingly poorer retention, are represented by the
mildly suppressive $\alpha=+0.5$ entry of Table~\ref{tab:astro_kernels}. The effective value
of $\alpha$ is not fixed uniquely by cluster calculations; these are phenomenological
representatives spanning the two behaviors.

\paragraph{Density versus kernel: why isolated channels are suppressive.}
Before assigning exponents to the isolated channels it is worth separating two effects that are
easily conflated, since the distinction recurs in a sharper form for PBHs in
Sec.~\ref{sec:pbh_conversion}. A statement that some environment ``produces more massive black
holes'' is a statement about the mass function $c(m,t)$, i.e.\ about the initial data of the
coagulation problem. The kernel answers a different question: given two black holes that already
exist, with masses $m_1$ and $m_2$, what is the per-pair probability per unit time that they
coalesce? These need not point the same way, and for isolated binary evolution they point in
opposite directions.

For dynamically assembled populations the kernel is set by encounter physics. Gravitational
focusing, mass segregation, and dynamical friction all concentrate the most massive objects in
the core and enhance their interaction cross-sections, so $K$ increases with mass and
$\lambda>0$. For isolated field evolution there are no encounters: the per-pair merger
probability is instead the probability that a given progenitor pair survives to leave a compact
binary tight enough to coalesce within a Hubble time. That probability \emph{decreases} with
progenitor mass. More massive progenitors reach larger radii on the giant branch, so mass
transfer begins at wider separations and the post-common-envelope orbit inherits a larger floor;
since the Peters time scales as $a_0^4$, the resulting coalescence times grow rapidly with mass.
The envelope binding energy also grows faster than the orbital energy available to unbind it, so
the required shrinkage becomes harder to achieve and common-envelope survival becomes less
likely at high mass~\cite{Dominik:2012kk, Belczynski:2016obo}. Both effects suppress $K$ at
large $m_1+m_2$, i.e.\ $\alpha>0$ and $\lambda<0$.

This gives Table~\ref{tab:astro_kernels} a coherent physical ordering rather than an arbitrary
span of exponents: \emph{dynamically assembled channels are mass-favoring
($\lambda>0$), while isolated binary-evolution channels are mass-suppressing ($\lambda<0$)},
with AGN disks intermediate. It is precisely because these two families sit on opposite sides of
$\lambda=0$ that the coagulation classification is informative about the environment.

\paragraph{Low-metallicity environments.}
In metal-poor or primordial environments, reduced stellar winds allow more massive BH
progenitors to form~\cite{Belczynski:2016obo}. As just noted, this enters the coagulation
problem through the initial mass function and not through the kernel. The kernel itself remains
suppressive for the reasons above, and low metallicity strengthens rather than weakens the
suppression: weaker winds leave more massive envelopes to be ejected during the common-envelope
phase, and the larger radii of metal-poor giants push the post-interaction separation further
out. We adopt $\alpha\sim1$, hence $\lambda\sim-1$, as a representative value. The precise
exponent depends on the adopted common-envelope efficiency and initial-separation distribution
and is not fixed uniquely by population-synthesis calculations.

\paragraph{Isolated field binaries.}
Low metallicity and binary-interaction physics can produce unusually massive first-generation
black-hole binaries in isolated evolution~\cite{Marchant:2016wow, Stevenson:2017tfq}, but such
calculations do not imply repeated hierarchical mergers. The $\alpha=2$ row of
Table~\ref{tab:astro_kernels} represents the limit in which the common-envelope and
giant-radius bottlenecks described above are strongest, so that the per-pair merger probability
falls steeply with total mass. It is the most suppressive case we consider and serves as the
opposite extreme to the nuclear-cluster entry; a population evolving under it coarsens slowly,
with $z=1/3$.

\paragraph{AGN disks.}
In gaseous active-galactic-nucleus disks, gas torques can drive compact objects toward
migration traps where encounters become more frequent~\cite{Bellovary:2016migration}; AGN-disk
population models consequently predict merger populations whose mass spectra can differ
substantially from isolated binaries~\cite{Yang:2019cbr, McKernan:2014agn}. The detailed mass
dependence depends on the disk model, so the $\alpha\simeq0$ entry in
Table~\ref{tab:astro_kernels} is an illustrative weak-mass-dependence limit.

\paragraph{Multi-channel composition.}
Real astrophysical populations involve several merger mechanisms. At the level of the collision
operator one may write $K_{\rm total}=\sum_i f_i(t)K_i$. Because the Smoluchowski operator is
linear in $K$ but quadratic in $c$, one has
$\mathcal C_{K_{\rm total}}[c]=\sum_i f_i\mathcal C_{K_i}[c]$, yet the solution for
$K_{\rm total}$ is \emph{not} the sum of the solutions obtained from the individual kernels. If
one channel dominates over a sufficiently long interval, its homogeneity can control an
intermediate asymptotic regime; when several channels are comparable, crossovers must be solved
for using the full Eq.~\eqref{eq:smoluchowski} rather than by superposing self-similar profiles.

%---------------------------------------------------------------------
\subsection{Mass-gap phenomenology}
%---------------------------------------------------------------------

The pair-instability supernova mechanism suppresses direct stellar-remnant BH formation in the
range $\sim50$--$120\,\msun$~\cite{Farmer:2019jed, Marchant:2018kun}. Hierarchical mergers can
populate this region by adding lower-mass progenitors, provided merger remnants are retained.
In the coagulation description the rate at which probability flows into the gap depends on the
full kernel and on the initial spectrum, not on the homogeneity exponent alone: kernels that
strongly suppress high-mass encounters produce a rapidly declining high-mass tail, whereas
kernels favoring massive participants populate the gap more efficiently. Events such as
GW190521~\cite{Abbott:2020mjq, Abbott:2020tfl} motivate hierarchical channels but do not select
a unique coagulation kernel. We note that a hard threshold, $K=0$ for $m+m'>C$, lies outside the
homogeneous class; in that toy problem probability accumulates near the upper active-mass
boundary, as a consequence of the imposed threshold rather than of the physical pair-instability
mechanism.

%---------------------------------------------------------------------
\subsection{Time dependence of merger rates}\label{sec:astro_time}
%---------------------------------------------------------------------

The factor $t^{-\delta}$ in Eq.~\eqref{eq:general_kernel} encodes the evolution of
the merger rate with cosmic time.  Two contributions dominate:
\begin{enumerate}
\item \textbf{Cosmic star formation history.}  The merger rate inherits the star
      formation rate $\dot\rho_\star(t_c)$ convolved with a delay-time distribution
      $P_{\rm delay}$:
      \begin{equation}
      S(t) = \int_0^{t} \dot\rho_\star(t_c)\,P_{\rm delay}(t-t_c)\,dt_c,
      \label{eq:DTD_convolution}
      \end{equation}
      where a broad delay-time distribution is commonly produced by binary evolution and
      GW inspiral~\cite{Dominik:2012kk}. Over a restricted redshift interval we approximate the
      resulting convolution by $S(t)\propto t^{-\delta}$; the range $\delta\sim0$--$1$ used below
      is a phenomenological local fit, not a universal consequence of the cosmic star-formation
      history~\cite{Mapelli:2019bpo}.
\item \textbf{Environment-specific dynamics.} In dense stellar clusters the central density
      evolves through relaxation, core contraction, and binary heating, modifying the encounter
      rate on cluster-evolution timescales. In AGN disks, finite disk lifetimes and gas-driven
      migration introduce additional time dependence~\cite{Bellovary:2016migration,
      Yang:2019cbr}.
\end{enumerate}
The separable factor $t^{-\delta}$ is most cleanly handled through the cumulative coagulation
clock $T(t)=\int_{t_i}^{t}t'^{-\delta}dt'$. For $\delta<1$, $T\propto t^{1-\delta}$ at late
times and the algebraic exponent in physical time is the $z$ derived in Sec.~\ref{sec:scaling}.
For $\delta=1$, $T\propto\ln t$ and the growth is logarithmic in physical time. For
$\delta>1$, $T(t)$ approaches a finite limit as $t\to\infty$: coagulation freezes
out rather than producing a characteristic mass that decreases with time. A negative value
obtained by naively inserting $\delta>1$ into the power-law formula should not be interpreted as
reverse coagulation.

\begin{table}[h]
\centering
\renewcommand{\arraystretch}{1.3}
\begin{tabular}{@{}lcccccc@{}}
\toprule
Environment & $\alpha$ & $\beta$ & $\delta$ & $\lambda$ & $z$ & $\theta$ \\
\midrule
Nuclear star cluster    & $-0.5$  & $0$    & $\approx0$ & $0.5$  & $2.00$ & $4.00$ \\
Globular cluster        & $0.5$   & $0$    & $\approx0$ & $-0.5$ & $0.67$ & $1.33$ \\
Low-metallicity field   & $1$     & $0$    & $\approx0$ & $-1$   & $0.50$ & $1.00$ \\
Isolated field binary   & $2$     & $0$    & $\approx0$ & $-2$   & $0.33$ & $0.67$ \\
AGN disk                & $\approx0$ & $0$ & $\approx0$ & $0$ & $1.00$ & $2.00$ \\
\bottomrule
\end{tabular}
\caption{Illustrative effective kernel parameters spanning qualitatively different
astrophysical regimes. These exponents are phenomenological model choices rather than unique
values derived from the cited environments. For $\lambda<1$ and $\delta<1$,
$z=(1-\delta)/(1-\lambda)$ and $\theta=2z$ in the conservative similarity regime.}
\label{tab:astro_kernels}
\end{table}

%=====================================================================
\section{Primordial Black Hole Merger Kernels}\label{sec:pbh}
%=====================================================================

Primordial black holes differ fundamentally from astrophysical BHs in their formation
mechanism and initial spatial distribution.  PBHs formed during the radiation-dominated
era from the collapse of enhanced primordial density
fluctuations~\cite{Carr:2016drx, Carr:2020gox, Sasaki:2018dmp}.  Their abundance,
quantified by $f_{\rm PBH}\equiv\Omega_{\rm PBH}/\Omega_{\rm DM}$, and their mass
function $\psi(m)$ set the initial conditions for the coagulation problem.

Binary formation proceeds through four distinct channels classified by epoch (early $E$
or late $L$) and multiplicity (two-body or three-body):
\begin{itemize}
\item \textbf{E2}: pairs decoupling from the Hubble flow in the early Universe;
\item \textbf{E3}: compact three-body configurations in the early Universe that eject
      one member to form a hard binary;
\item \textbf{L2}: two-body gravitational-wave capture during late-time halo encounters;
\item \textbf{L3}: three-body hardening encounters in dense late-time halos.
\end{itemize}
We adopt the standard notation $M\equiv m_1+m_2$, $\mu\equiv m_1 m_2/M$,
$\eta\equiv\mu/M = m_1 m_2/M^2$, and the normalized mass function
\begin{equation}
\psi(m)\equiv\frac{m}{\rho_{\rm PBH}}\frac{dn_{\rm PBH}}{d\ln m},
\qquad
\int\psi(m)\,d\ln m=1,
\qquad
\langle m\rangle=\frac{\rho_{\rm PBH}}{n_{\rm PBH}}.
\label{eq:psi_def}
\end{equation}
Note for later use that Eq.~\eqref{eq:psi_def} relates $\psi$ directly to the coagulation
variable $c(m,t)=dn_{\rm PBH}/dm$:
\begin{equation}
\psi(m)=\frac{m^{2}\,c(m,t)}{\rho_{\rm PBH}}.
\label{eq:psi_to_c}
\end{equation}

%---------------------------------------------------------------------
\subsection{Physical prerequisites}
%---------------------------------------------------------------------

In geometric units ($c=G=1$), a Keplerian binary with component masses $m_1,m_2$,
semi-major axis $r_a$, and dimensionless angular momentum
$j\equiv (\mathcal{L}/\mu)/\sqrt{r_a M}$ has eccentricity $e=\sqrt{1-j^2}$, where
$\mathcal{L}$ is the orbital angular momentum.
For the highly eccentric orbits that dominate early-Universe formation ($j\ll 1$), the Peters
coalescence time due to GW emission is~\cite{Peters:1964, Nakamura:1997sm}
\begin{equation}
\tau_{\rm coal} = \frac{3}{85}\frac{r_a^4\,j^7}{\eta\,M^3},
\label{eq:coal_time}
\end{equation}
which has an extreme sensitivity to $j$.  This sensitivity is the reason that the
angular-momentum distribution at formation is the key uncertainty in all PBH merger
rate calculations~\cite{Ali-Haimoud:2017rtz, Raidal:2024bmm}.

%---------------------------------------------------------------------
\subsection{Early two-body (E2)}
%---------------------------------------------------------------------

\paragraph{Formation mechanism.}
A pair of PBHs of masses $m_1, m_2$ at comoving separation $x_0$ decouples from the
Hubble flow when their mutual overdensity exceeds unity.  In the absence of a third
nearby PBH within comoving radius $y$, the pair probability follows a Poisson
distribution~\cite{Nakamura:1997sm, Sasaki:2016jop}:
\begin{equation}
dn_{\rm pairs} =
\frac{1}{2}e^{-\bar{N}(y)}\,dn(m_1)\,dn(m_2)\,dV(x_0),
\label{eq:pair_prob}
\end{equation}
with $\bar{N}(y)=n\,V(y)$ the expected number of PBHs inside radius $y$.  The pair
decouples at scale factor $a_{\rm dc} = a_{\rm eq}/\delta_{\rm pair}$, where
$\delta_{\rm pair}=(M/2)/[\rho_M V(x_0)]$, and the resulting semi-major axis is
$r_a\approx 0.1\,a_{\rm dc}x_0 \approx 0.84\,\rho_R x_0^4/M$.

\paragraph{Angular momentum.}
Tidal torques from surrounding matter source the orbital angular momentum:
$j = j_{\rm PBH}+j_M$, where the PBH contribution dominates at small
$\bar{N}(x_0)$.  The reference scale is~\cite{Ali-Haimoud:2017rtz}
\begin{equation}
j_0\equiv 0.95\,\bar{N}(x_0)\,\frac{\langle m\rangle}{M}
\approx 0.4\,\frac{f_{\rm PBH}}{\delta_{\rm pair}},
\label{eq:j0_E2}
\end{equation}
giving $j_0\approx 10^{-2}$ for binaries merging today (highly eccentric).  The angular
momentum distribution is
\begin{equation}
j\frac{dP}{dj} =
\frac{j^2/j_0^2}{\bigl(1+j^2/j_0^2\bigr)^{3/2}}
\quad [\bar{N}(y)\ll 1],
\label{eq:jdist_E2}
\end{equation}
transitioning to a Gaussian $j\,dP/dj\propto j^2\exp(-j^2/\sigma_j^2)$ when
$\bar{N}(y)\gg 1$.

\paragraph{Merger rate.}
Accounting for binary disruption at formation ($S_E$) and inside dark matter halos
($S_L$)~\cite{Ali-Haimoud:2017rtz, Raidal:2024bmm},
$dR_{\rm E2}/dm_1dm_2 = S_LS_E\,dR^{(0)}_{\rm E2}/dm_1dm_2$, where the unsuppressed rate is
\begin{equation}
\frac{dR^{(0)}_{\rm E2}}{dm_1\,dm_2}
\approx
\frac{1.6\times 10^6}{{\rm Gpc}^3\,{\rm yr}}\,
f_{\rm PBH}^{53/37}\,
\eta^{-34/37}\,
\left(\frac{M}{\msun}\right)^{-32/37}
\left(\frac{t}{t_0}\right)^{-34/37}
\frac{\psi(m_1)\psi(m_2)}{m_1 m_2},
\label{eq:RE2_unif}
\end{equation}
and the late-time suppression factor, valid for $f_{\rm PBH}\gtrsim 0.1$, is
\begin{equation}
S_L(t)\approx
\min\!\left(1,\;
0.01
\left[\left(\frac{t}{t_0}\right)^{0.44}\!f_{\rm PBH}\right]^{-0.65}
\!\!\exp\!\left\{
0.03\ln^2\!\left[\left(\frac{t}{t_0}\right)^{0.44}\!f_{\rm PBH}\right]
\right\}\right).
\label{eq:SL_E2}
\end{equation}
The exponents of the \emph{rate density} are $a'=32/37$, $b'=34/37$, and $\delta=34/37$;
Sec.~\ref{sec:pbh_conversion} converts them into kernel exponents.

%---------------------------------------------------------------------
\subsection{Early three-body (E3)}
%---------------------------------------------------------------------

The E3 channel becomes significant when $f_{\rm PBH}\gtrsim 0.1$ or when PBHs form in
clusters~\cite{Raidal:2024bmm}.  Three PBHs with masses $m_1, m_2, m_3$ undergo
hierarchical decoupling: the $m_1$--$m_2$ pair at comoving separation $x$ decouples
first, while the third PBH at distance $y > x$ decouples later, imparting angular
momentum
\begin{equation}
j_{12} = 1.4\,\frac{m_3}{m_1+m_2}\,\frac{\bar{N}(x)}{\bar{N}(y)}\,|\sin 2\vartheta|,
\label{eq:j12_E3}
\end{equation}
where $\vartheta$ is the angle between the separation vectors.  The post-interaction
angular-momentum distribution is $dP(j)/dj = \gamma\,j^{\gamma-1}$ with $\gamma\in[1,2]$
($\gamma=1$ from $N$-body simulations, $\gamma=2$ for a thermal distribution).

The merger rate is~\cite{Raidal:2024bmm}
\begin{align}
\frac{dR_{\rm E3}}{d\ln m_1\,d\ln m_2}
&= R_{\rm E3,mono}
\left(\frac{M}{2\langle m\rangle}\right)^{\!\frac{179\gamma}{259}-\frac{2122}{333}}
(4\eta)^{-\frac{3\gamma}{7}-1}
\,\overline{\mathcal{F}}(m_1,m_2)\,\psi(m_1)\psi(m_2),
\label{eq:RE3}
\end{align}
where, for a monochromatic mass function,
\begin{align}
R_{\rm E3,mono} =
\frac{7.9\times 10^4}{{\rm Gpc}^3\,{\rm yr}}\,
f_{\rm PBH}^{\frac{144\gamma}{259}+\frac{47}{37}}
\left[\frac{t}{t_0}\right]^{\frac{\gamma}{7}-1}
\!\left[\frac{\langle m\rangle}{\msun}\right]^{\frac{5\gamma-32}{37}}
\frac{e^{3.2(1-\gamma)}\,\gamma}{28/9-\gamma}\,\mathcal{K},
\label{eq:RE3mono}
\end{align}
with $\mathcal{K}\approx 4.0$ encoding three-body dynamics.

%---------------------------------------------------------------------
\subsection{Late two-body (L2)}
%---------------------------------------------------------------------

Two PBHs on initially unbound trajectories can become bound when a close passage radiates
more energy in gravitational waves than their kinetic energy at infinity. At leading
quadrupole order, gravitational focusing gives the maximum impact parameter
\begin{equation}
b_{\rm max}=\left(\frac{340\pi}{3}\right)^{1/7}
\frac{GM}{c^2}\,\eta^{1/7}\left(\frac{v_{\rm rel}}{c}\right)^{-9/7},
\label{eq:gw_capture_bmax}
\end{equation}
with $M=m_1+m_2$ and $\eta=m_1m_2/M^2$~\cite{Peters:1964, Mouri:2002}. Averaging
over a Maxwell-Boltzmann velocity distribution with one-dimensional dispersion $\sigma_v$
and integrating over an effective overdensity $\delta_{\rm eff}$ relative to the cosmic
mean gives the population merger-rate expression of Ref.~\cite{Raidal:2024bmm},
\begin{align}
\frac{dR_{\rm L2}}{d\ln m_1\,d\ln m_2}
\approx
\frac{3.4\times 10^{-6}}{{\rm Gpc}^3\,{\rm yr}}\,
f_{\rm PBH}^2\,\delta_{\rm eff}
\left(\frac{\sigma_v}{\kms}\right)^{-11/7}
\eta^{-5/7}\,\psi(m_1)\psi(m_2).
\label{eq:RL2}
\end{align}
This channel is time-independent at fixed $\delta_{\rm eff}$, i.e.\ $\delta=0$, with rate-density
exponents $a'=0$, $b'=5/7$.

It is worth noting that the L2 kernel can be obtained directly from
Eq.~\eqref{eq:gw_capture_bmax} without reference to Eq.~\eqref{eq:RL2}, and this provides an
independent check on the conversion of Sec.~\ref{sec:pbh_conversion}. The capture cross-section
is $\sigma=\pi b_{\rm max}^2\propto M^{2}\eta^{2/7}v_{\rm rel}^{-18/7}$, so the per-pair rate
coefficient is
\begin{equation}
K_{\rm L2}=\langle\sigma v_{\rm rel}\rangle
\propto M^{2}\,\eta^{2/7}\,\sigma_v^{-11/7},
\label{eq:KL2_direct}
\end{equation}
i.e.\ $a=-2$, $b=-2/7$, and $\lambda=+2$.

%---------------------------------------------------------------------
\subsection{Late three-body (L3)}
%---------------------------------------------------------------------

In sufficiently dense halos a third PBH extracts energy from a two-body encounter via
the Heggie--Hills mechanism, leaving a hard bound binary~\cite{Raidal:2024bmm}.  The
hardness factor $\mathcal{H} \equiv |E_{\rm bin}|/\langle KE\rangle$ at formation must exceed
a minimum value $\mathcal{H}_{\rm min}\simeq 5$ for the binary to survive subsequent
encounters.
With the angular-momentum distribution $dP/dj = \gamma j^{\gamma-1}$ (same
parametrization as E3), the merger rate is
\begin{align}
\frac{dR_{\rm L3}}{d\ln m_1\,d\ln m_2}
&\approx
\frac{1.3\times 10^{-16}\,e^{-6.0(\gamma-1)}}{{\rm Gpc}^3\,{\rm yr}}\,
f_{\rm PBH}^3\,\delta_{\rm eff}^2
\left(\frac{\sigma_v}{\kms}\right)^{-9+\frac{8\gamma}{7}}
\nonumber\\
&\quad\times
\eta^{-1+\frac{\gamma}{7}}
\left(\frac{M}{\msun}\right)^{3-\frac{\gamma}{7}}
\left(\frac{t}{t_0}\right)^{\frac{\gamma}{7}}
\mathcal{F}\!\left(\frac{\langle m\rangle\,\mathcal{H}_{\rm min}}{2\eta M}\right)
\psi(m_1)\psi(m_2),
\label{eq:RL3}
\end{align}
with $\mathcal{H}_{\rm min}\simeq 5$.  Unlike L2, this channel has a weak power-law time
dependence: $\delta = -\gamma/7 < 0$ for $\gamma > 0$, meaning the L3 rate grows with
cosmic time. Since $\delta<0$ implies that the coagulation clock $T(t)$ grows faster than
linearly, this channel accumulates coagulation time more rapidly than any other, which is
relevant below when assessing whether a formally gelling kernel reaches its gel point.

%---------------------------------------------------------------------
\subsection{From merger rates to coagulation kernels}
\label{sec:pbh_conversion}
%---------------------------------------------------------------------

PBH merger rates are conventionally quoted as rate densities in the $M$--$\eta$
parametrization,
\begin{equation}
\frac{dR}{d\ln m_1\,d\ln m_2}
= C\,M^{-a'}\,\eta^{-b'}\,t^{-\delta}\,\psi(m_1)\,\psi(m_2),
\label{eq:pbh_meta_kernel}
\end{equation}
and it is tempting to insert $(a',b')$ directly as $(\alpha,\beta)$ in
Eq.~\eqref{eq:general_kernel}. This is incorrect for two independent reasons, and both must
be undone before the coagulation formalism can be applied.

First, $\eta$ and $M$ are not the arguments of Eq.~\eqref{eq:general_kernel}. Since
$M^{-a}\eta^{-b}=M^{-a+2b}(m_1m_2)^{-b}$, matching onto
$K\propto M^{-\alpha}(m_1m_2)^{-\beta}$ gives
\begin{equation}
\alpha=a-2b,\qquad \beta=b,\qquad \lambda=-(\alpha+2\beta)=-a.
\label{eq:pbh_conversion}
\end{equation}

Second, and more consequentially, Eq.~\eqref{eq:pbh_meta_kernel} is a rate \emph{density},
whereas $K$ is the coefficient of $c(m_1)c(m_2)$ in Eq.~\eqref{eq:rate_kernel_relation}. Using
Eq.~\eqref{eq:psi_to_c},
\begin{equation}
\frac{dR}{d\ln m_1\,d\ln m_2}
= C\,M^{-a'}\eta^{-b'}t^{-\delta}\,\frac{m_1^2m_2^2}{\rho_{\rm PBH}^2}\,c(m_1)c(m_2),
\end{equation}
so that, converting to $dR/dm_1dm_2$ and comparing with
Eq.~\eqref{eq:rate_kernel_relation},
\begin{equation}
K \;\propto\; M^{-a'}\eta^{-b'}t^{-\delta}\times m_1m_2
\;=\; M^{-(a'-2)}\,\eta^{-(b'-1)}\,t^{-\delta},
\label{eq:rate_to_kernel}
\end{equation}
where we used $m_1m_2=\eta M^{2}$. The kernel exponents are therefore
\begin{equation}
\boxed{a=a'-2,\qquad b=b'-1,\qquad \lambda=-a=2-a'.}
\label{eq:pbh_ab_shift}
\end{equation}
Note that $\alpha=a-2b$ is invariant under this shift, so only $\beta$ and $\lambda$ are
affected. For a nongelling channel with $\lambda<1$ and $\delta<1$,
\begin{equation}
z=\frac{1-\delta}{1+a},\qquad \theta=2z.
\label{eq:pbh_z}
\end{equation}

Equation~\eqref{eq:KL2_direct} provides an independent check: computing $\langle\sigma v\rangle$
directly from the capture cross-section gives $a=-2$, $b=-2/7$ for L2, exactly as
Eq.~\eqref{eq:pbh_ab_shift} returns from $a'=0$, $b'=5/7$. A second check is provided by the E2
channel in the monochromatic limit, where the rate density integrates to
$R\propto f_{\rm PBH}^{53/37}m^{-32/37}$ while $n=f_{\rm PBH}\rho_{\rm DM}/m$, so that
$K=2R/n^2\propto m^{42/37}$ and $\lambda_{\rm E2}=42/37$, again matching
Eq.~\eqref{eq:pbh_ab_shift}.

Tables~\ref{tab:pbh_kernel_exponents} and~\ref{tab:scaling_solutions} collect the rate-density
exponents and the resulting kernel classification.

\begin{table}[h]
\centering
\renewcommand{\arraystretch}{1.3}
\resizebox{\textwidth}{!}{%
\begin{tabular}{@{}lcccc@{}}
\toprule
Pathway & $a'$ in $M^{-a'}$ & $b'$ in $\eta^{-b'}$ & $\delta$ & Dominant when \\
\midrule
E2 & $32/37 \simeq 0.865$ & $34/37 \simeq 0.919$ & $34/37 \simeq 0.919$ & $f_{\rm PBH}\ll1$ \\
E3 ($\gamma=1$) & $2122/333-179/259 \simeq 5.681$ & $10/7 \simeq 1.429$ & $6/7 \simeq 0.857$ & $f_{\rm PBH}\gtrsim0.1$ \\
E3 ($\gamma=2$) & $2122/333-358/259 \simeq 4.990$ & $13/7 \simeq 1.857$ & $5/7 \simeq 0.714$ & $f_{\rm PBH}\gtrsim0.1$ \\
L2 & $0$ & $5/7 \simeq 0.714$ & $0$ & late-time halos \\
L3 ($\gamma=1$) & $-20/7 \simeq -2.857$ & $6/7 \simeq0.857$ & $-1/7 \simeq-0.143$ & dense halos \\
L3 ($\gamma=2$) & $-19/7 \simeq -2.714$ & $5/7 \simeq0.714$ & $-2/7 \simeq-0.286$ & dense halos \\
\bottomrule
\end{tabular}%
}
\caption{PBH merger-rate \emph{density} exponents in the native $M$--$\eta$ parametrization
of Eq.~\eqref{eq:pbh_meta_kernel}. The primed symbols $a'$ and $b'$ are deliberately kept
distinct from the kernel exponents $a$, $b$ of Eq.~\eqref{eq:pbh_ab_shift} and from the
master-kernel parameters $\alpha$, $\beta$.}
\label{tab:pbh_kernel_exponents}
\end{table}

\begin{table}[h]
\centering
\renewcommand{\arraystretch}{1.25}
\resizebox{\textwidth}{!}{%
\begin{tabular}{@{}lrrrrrcl@{}}
\toprule
Pathway & $a=a'-2$ & $b=b'-1$ & $\alpha=a-2b$ & $\beta=b$ & $\lambda=-a$ & $\delta$ & Regime \\
\midrule
E2 & $-1.135$ & $-0.081$ & $-0.973$ & $-0.081$ & $+1.135$ & $0.919$ & gelling candidate \\
E3 ($\gamma=1$) & $3.681$ & $0.429$ & $2.824$ & $0.429$ & $-3.681$ & $0.857$ & nongelling, $z=0.0305$ \\
E3 ($\gamma=2$) & $2.990$ & $0.857$ & $1.276$ & $0.857$ & $-2.990$ & $0.714$ & nongelling, $z=0.0716$ \\
L2 & $-2.000$ & $-0.286$ & $-1.429$ & $-0.286$ & $+2.000$ & $0$ & gelling candidate \\
L3 ($\gamma=1$) & $-4.857$ & $-0.143$ & $-4.571$ & $-0.143$ & $+4.857$ & $-0.143$ & gelling candidate \\
L3 ($\gamma=2$) & $-4.714$ & $-0.286$ & $-4.143$ & $-0.286$ & $+4.714$ & $-0.286$ & gelling candidate \\
\bottomrule
\end{tabular}%
}
\caption{PBH channels converted to the master-kernel convention of
Eq.~\eqref{eq:general_kernel} using Eq.~\eqref{eq:pbh_ab_shift}. Only the early three-body
channel has $\lambda<1$ and admits the mass-conserving power-law exponent of
Eq.~\eqref{eq:pbh_z}. The E2, L2 and L3 rates all give $\lambda>1$ and therefore cannot be
assigned a mass-conserving coarsening exponent; they must instead be analyzed as candidate
gelling channels with their coagulation clocks taken into account.}
\label{tab:scaling_solutions}
\end{table}

Three caveats attach to this mapping. First, the E2, E3 and L3 rates depend on population-level
quantities---$f_{\rm PBH}$, $\langle m\rangle$, $\delta_{\rm eff}$---that are functionals of the
whole distribution and therefore evolve during coagulation. Strictly, these are not pure
two-body kernels, and the mean-field mapping is approximate in a way that the homogeneity
classification does not capture. Second, the quoted $t^{-\delta}$ factors are not per-pair rate
coefficients: the E2 factor $t^{-34/37}$, for instance, arises from the distribution of
coalescence delay times of binaries formed at $z\sim10^{10}$, not from an environmental
modulation of an instantaneous pairing rate. Inserting it into the coagulation clock therefore
double-counts the merger timing to some degree. We retain it for definiteness, but the
resulting $z$ values for the nongelling channel should be read as order-of-magnitude estimates.
Third, the E3 and L3 rates carry additional factors $\overline{\mathcal{F}}(m_1,m_2)$ and
$\mathcal{F}(\langle m\rangle\mathcal{H}_{\rm min}/2\eta M)$ that we do not expand. The
classification of Table~\ref{tab:scaling_solutions} is therefore a classification of the
leading $M$--$\eta$ power dependence of each rate, not a proof that these additional factors
are mass-independent; if they carry residual homogeneity of their own it would shift the
tabulated $\lambda$ values, and we have not verified that they do not.

%---------------------------------------------------------------------
\subsection{Channel dominance and comparative analysis}
\label{sec:pbh_dominance}
%---------------------------------------------------------------------

The relative importance of the four channels depends on $f_{\rm PBH}$, the clustering
of the PBH population, and the epoch.

\paragraph{Small $f_{\rm PBH}$ ($\lesssim 0.01$).}
The E2 channel dominates overwhelmingly.  PBHs are sparsely distributed, three-body
configurations are rare, and late-time halo densities are low.  The total merger rate
scales as $\mathcal{R} \propto f_{\rm PBH}^{53/37}$ and is suppressed by the
disruption factor $S_L \lesssim 1$ from halo encounters~\cite{Raidal:2024bmm}.

\paragraph{Intermediate $f_{\rm PBH}$ ($0.01 \lesssim f_{\rm PBH} \lesssim 0.1$).}
The E2 rate still dominates but is increasingly suppressed by $S_E$ (disruption during
the radiation-dominated era) and $S_L$ (disruption in halos).  The L2 channel, which
scales as $f_{\rm PBH}^2 \delta_{\rm eff}$, begins to contribute comparably in dense
halo environments at late times.

\paragraph{Large $f_{\rm PBH}$ ($\gtrsim 0.1$).}
Early three-body configurations become frequent.  The E3 rate scales as
$f_{\rm PBH}^{144\gamma/259 + 47/37}$, which grows faster with $f_{\rm PBH}$ than
E2 at large $f_{\rm PBH}$.  Simultaneously, both late channels are enhanced by the higher halo
densities.  For $f_{\rm PBH}\approx1$ several channels contribute simultaneously; the effective
kernel is then a sum of channel kernels, but the resulting mass function must be evolved with
the full combined collision operator and is not a linear superposition of the single-channel
similarity profiles.

\paragraph{Clustering boost.}
If PBHs form in clusters (Poisson deviations in the initial power spectrum), the effective
local $f_{\rm PBH}$ inside halos is enhanced by a factor $\delta_{\rm eff} \gg 1$,
boosting the L2 and L3 rates~\cite{Raidal:2024bmm, Bird:2016dcv}.  The E2 and E3 rates,
fixed at formation in the early Universe, are not directly affected by late-time
clustering but are modified through the suppression factor $S_L$.

\paragraph{Observational implications.}
For standard early-binary assumptions and sufficiently narrow stellar-mass PBH mass
functions, the E2 channel can yield merger rates of the same order as LVK-inferred BBH rates
for $f_{\rm PBH}$ in the approximate $10^{-3}$--$10^{-2}$ range, though the abundance inference
is model dependent through the mass function and suppression
factors~\cite{Ali-Haimoud:2017rtz, LIGOScientific:2021djp}. The corrected classification of
Table~\ref{tab:scaling_solutions} carries a qualitative implication for this comparison. Because
$\lambda_{\rm E2}=42/37>1$, the E2 kernel weights heavy pairs superlinearly: in coagulation
language, this is the statement---familiar from direct calculations of PBH pairing with extended
mass functions---that the early-binary merger rate is dominated by the heaviest objects present.
A mass-conserving self-similar description with a slowly growing characteristic mass is
therefore \emph{not} the appropriate idealization for E2, and the assumption that hierarchical
mergers leave the PBH mass function essentially unmodified requires justification from the full
kernel, including the suppression factors $S_E$ and $S_L$ and the finite coagulation clock,
rather than from a homogeneity exponent alone. We emphasize that $\lambda>1$ is a diagnostic and
not a theorem: whether a physical runaway develops depends on the behavior of the kernel at very
unequal masses and on whether $T(t)$ reaches the gel time of the autonomous problem, neither of
which is settled here.

%=====================================================================
\section{Self-Similar Analysis and the Scaling Function $\Phi$}
\label{sec:scaling}
%=====================================================================

Having classified the merger kernels in Secs.~\ref{sec:astrobh}--\ref{sec:pbh}, we now derive
the conservative long-time scaling structure of Eq.~\eqref{eq:smoluchowski}. The results in
this section assume $L=1$ and no injection or mass sink.

%---------------------------------------------------------------------
\subsection{General scaling form and the roles of $\theta$, $\tau$, $z$}
\label{sec:theta_tau_derivation}
%---------------------------------------------------------------------

A convenient parametrization of a self-similar solution is
\begin{equation}
c(m,t) \sim t^{-\theta}\,m^{-\tau}\,\Phi\!\left(\frac{m}{t^z}\right),
\qquad \xi \equiv m/t^z,
\label{eq:general_scaling_form}
\end{equation}
where $z$ controls the growth of a characteristic mass scale $s(t)\sim t^z$, $\theta$
controls the overall amplitude decay, and $m^{-\tau}$ allows for a possible algebraic
singularity at small masses. At fixed $\xi$ one has
$c(m,t) = t^{-(\theta+z\tau)}\xi^{-\tau}\Phi(\xi)$, so $\theta+z\tau$ is the effective time
exponent.

The factor $m^{-\tau}=t^{-z\tau}\xi^{-\tau}$ can always be absorbed into a redefined shape
function, so $(\theta,\tau)$ are not separately invariant labels of a similarity solution.
Mass conservation gives
$M_1(t)=t^{-\theta+z(2-\tau)}\int_{m_0/t^z}^{\infty}\xi^{1-\tau}\Phi(\xi)\,d\xi$ and hence,
when the similarity moment converges,
\begin{equation}
\theta+z\tau=2z.
\label{eq:combined_theta_relation}
\end{equation}
It is therefore convenient to choose the standard representation $\tau=0$ and define a single
scaling profile that contains any endpoint power law:
\begin{equation}
\boxed{c(m,t)=s(t)^{-2}\,\Phi\!\left(\frac{m}{s(t)}\right)}.
\label{eq:standard_similarity_form}
\end{equation}
In a physical-time power-law regime with $s\propto t^z$, this convention gives $\theta=2z$. The
low-mass cutoff ensures that no physical support is created below $m_0$, but it does not by
itself prove that $\Phi(\xi)$ is bounded as $\xi\to0$; endpoint behavior must be obtained from
the full similarity equation. The number density scales as
$N(t)=s(t)^{-1}\int_{m_0/s(t)}^{\infty}\Phi(\xi)\,d\xi$, provided the zeroth similarity moment
is finite.

%---------------------------------------------------------------------
\subsection{Coagulation clock and characteristic mass growth}
\label{sec:z_check}
%---------------------------------------------------------------------

Because the time dependence of Eq.~\eqref{eq:general_kernel} is separable, define the
cumulative coagulation clock
\begin{equation}
T(t)\equiv\int_{t_i}^{t}t'^{-\delta}\,dt'.
\label{eq:coag_clock}
\end{equation}
In terms of $T$, the conservative equation is the autonomous Smoluchowski equation with
mass-homogeneous degree $\lambda=-(\alpha+2\beta)$. In a nongelling scaling regime with
$\lambda<1$, dimensional balance gives $s(T)\propto T^{1/(1-\lambda)}$. For $\delta<1$,
$T\propto t^{1-\delta}$ at late times and therefore
\begin{equation}
\boxed{z=\frac{1-\delta}{1-\lambda}=\frac{1-\delta}{1+\alpha+2\beta},}
\qquad
c(m,t)\sim t^{-2z}\Phi\!\left(\frac{m}{t^z}\right).
\label{eq:z_formula}
\end{equation}
For $\delta=1$ this becomes $s\propto(\ln t)^{1/(1-\lambda)}$ rather than a power law in $t$;
for $\delta>1$, $T(t)$ approaches a finite limit and coagulation freezes out asymptotically. At
$\lambda=1$ the algebraic law is singular and marginal kernels require separate analysis; for
$\lambda>1$ the mass-conserving similarity form can fail through gelation or runaway growth.
These regimes are discussed further in Sec.~\ref{sec:regimes}.

%---------------------------------------------------------------------
\subsection{Similarity equation for $\Phi$}
\label{sec:phi_equation}
%---------------------------------------------------------------------

For a kernel homogeneous of degree $\lambda$ we define the dimensionless kernel $\kappa$ by
\begin{equation}
K(m,m',t)=t^{-\delta}\,s(t)^{\lambda}\,\kappa\!\left(\frac{m}{s},\frac{m'}{s}\right),
\qquad
\kappa(\xi,\zeta)=(\xi+\zeta)^{-\alpha}(\xi\zeta)^{-\beta},
\label{eq:kappa_def}
\end{equation}
so that $\kappa$ is the mass-rescaled kernel evaluated on similarity variables. Substituting
Eq.~\eqref{eq:standard_similarity_form} into Eq.~\eqref{eq:smoluchowski} and using
Eq.~\eqref{eq:z_formula}, the derivative term becomes
$\partial_tc=-(\dot s/s^{3})[2\Phi(\xi)+\xi\Phi'(\xi)]$ while the collision terms both carry
$s^{\lambda-3}$; the explicit time dependence therefore cancels provided
$w\equiv\dot s\,s^{-\lambda}$ is constant, which is precisely the condition
$s\propto T^{1/(1-\lambda)}$. The profile obeys the closed nonlinear integro-differential
equation
\begin{align}
w\!\left(2\Phi(\xi) + \xi\Phi'(\xi)\right)
&= \Phi(\xi)\int_0^\infty \kappa(\xi,\zeta)\,\Phi(\zeta)\,d\zeta
\nonumber\\
&\quad
- \frac{1}{2}\int_0^\xi \kappa(\xi-\zeta,\zeta)\,\Phi(\xi-\zeta)\,\Phi(\zeta)\,d\zeta,
\label{eq:phi_similarity_equation}
\end{align}
subject to $\int_0^\infty\xi\,\Phi(\xi)\,d\xi = M_1$ and
$\int_0^\infty\Phi(\xi)\,d\xi < \infty$. Equation~\eqref{eq:phi_similarity_equation} is the
central object of our analysis: it determines the late-time mass spectrum for given
$(\alpha,\beta)$.

It is important that the coefficient on the left is $w=\dot s\,s^{-\lambda}$ and not $z$
itself. Writing $s=(Ct)^{z}$ gives $w=zC$, so the amplitude of $\Phi$ and the normalization of
the characteristic mass are not independent: fixing one fixes the other. This is made explicit
by the constant-kernel solution of Sec.~\ref{subsec:constant-kernel-exact}.

\subsubsection{Endpoint behavior ($\xi\to0$)}
\label{sec:phi_endpoint}

The low-$\xi$ behavior cannot in general be inferred from the gain integral alone. If
$\Phi(\xi)\sim C\xi^{\nu}$, the loss term contains
\begin{equation}
\Phi(\xi)\int_0^{\infty}\kappa(\xi,\zeta)\Phi(\zeta)\,d\zeta
\sim C\,\xi^{\nu-\beta}\int_0^{\infty}\zeta^{-\alpha-\beta}\Phi(\zeta)\,d\zeta,
\label{eq:loss_small_xi_scaling}
\end{equation}
whenever the displayed moment exists, while the gain term samples the distinct region
$0<\zeta<\xi$ and, for a local power-law ansatz, scales as
\begin{equation}
\mathcal G(\xi)\sim \frac{C^2}{2}\,\xi^{1-\alpha-2\beta+2\nu}
\int_0^1[u(1-u)]^{\nu-\beta}\,du.
\label{eq:gain_small_xi_scaling}
\end{equation}
Convergence of the gain integral is therefore not by itself sufficient to establish a bounded
endpoint: the loss term may be more singular, and nonlocal contributions can change the
balance. In the numerical work the physical cutoff is retained at $\xi_{\min}=m_0/s(t)$ and the
endpoint exponent is measured from the converged solution.

\subsubsection{Large-$\xi$ tail and global fitting form}
\label{sec:phi_tail}

Mass conservation requires $\Phi(\xi)=o(\xi^{-2})$ if the first similarity moment is to be
finite. Many homogeneous nongelling kernels possess exponential or stretched-exponential tails
with an algebraic prefactor, motivating the flexible representation
\begin{equation}
\Phi(\xi)=A\left(\frac{\xi}{\xi_0}\right)^p
\exp\!\left[-\left(\frac{\xi}{\xi_0}\right)^q\right],
\label{eq:phi_interpolating}
\end{equation}
which we use throughout as an empirical descriptor. Homogeneity alone does not determine $q$:
its value depends on the detailed kernel, including its behavior for strongly unequal masses,
and no relation $q=q(\lambda)$ has been established for the present kernel class. The constant
kernel provides the exact benchmark $p=0$, $q=1$. The parameters $(A,\xi_0,p,q)$ are
phenomenological descriptors of the numerical solution; in particular, $p$ should not be
identified with the redundant prefactor exponent $\tau$ of
Eq.~\eqref{eq:general_scaling_form}.

\subsubsection{Classical solvable kernels}
For the additive kernel $K=m^{\lambda}+m'^{\lambda}$ one has
$\kappa(\xi,\zeta)=\xi^{\lambda}+\zeta^{\lambda}$; for the multiplicative kernel
$K=(mm')^{\lambda/2}$ one has $\kappa(\xi,\zeta)=(\xi\zeta)^{\lambda/2}$. In both cases
Eq.~\eqref{eq:z_formula} applies unchanged, since $1+\alpha+2\beta=1-\lambda$ for any kernel
of homogeneity $\lambda$. These classical families are distinct from the suppressive class
studied here: they favor mergers of large masses ($\lambda>0$), whereas
Eq.~\eqref{eq:general_kernel} with $\alpha,\beta>0$ suppresses them ($\lambda<0$).

%=====================================================================
\subsection{Exact constant-kernel benchmark}
\label{subsec:constant-kernel-exact}

The simplest member of the kernel class is $(\alpha,\beta,\delta)=(0,0,0)$, i.e.\ $K=1$,
with $\lambda=0$ and hence $z=1$, $\theta=2$: the characteristic mass grows linearly in time
and the amplitude decays as $t^{-2}$.

For definiteness, consider monodisperse initial data $c(m,0)=n_0\,\delta(m-m_0)$, where $n_0$
is the initial number density and $m_0$ the seed mass. Since mergers only add masses, the exact
solution remains discrete, with allowed masses $m=km_0$, $k=1,2,\ldots$. Writing $c_k(t)$ for
the number density of objects of mass $km_0$, the discrete Smoluchowski equation is
\begin{equation}
\frac{d c_k}{dt}
=\frac{1}{2}\sum_{i+j=k}c_i c_j-c_k N(t),
\qquad
N(t)\equiv \sum_{k=1}^{\infty}c_k(t),
\end{equation}
and summing over $k$ gives $dN/dt=-N^2/2$, so that $N(t)=n_0/(1+\hat t\,)$ with
$\hat t\equiv n_0t/2$.
The exact mass spectrum is the geometric distribution
\begin{equation}
c_k(t)=\frac{n_0}{(1+\hat t\,)^2}\left(\frac{\hat t}{1+\hat t}\right)^{k-1},
\qquad k=1,2,\ldots,
\label{eq:constant_kernel_discrete_solution}
\end{equation}
which conserves total mass exactly, $\sum_k km_0c_k = n_0m_0$, while the mean generation index
grows as $\langle k\rangle=1+\hat t$. The characteristic physical mass is therefore
\begin{equation}
s(t)=m_0(1+\hat t\,)=m_0\left(1+\frac{n_0t}{2}\right)
\;\xrightarrow[\ \hat t\gg1\ ]{}\;
\frac{M_1\,t}{2},
\qquad M_1=n_0m_0.
\label{eq:constant_kernel_s}
\end{equation}
At late times, for masses $m=km_0$ with $k\sim\hat t$, the discrete solution approaches a
continuous self-similar form. With $\xi\equiv m/s(t)\simeq k/\hat t$,
Eq.~\eqref{eq:constant_kernel_discrete_solution} gives
$c_k\simeq (n_0/\hat t^{\,2})\exp(-k/\hat t\,)$, i.e.
\begin{equation}
c(m,t)=\frac{1}{s(t)^{2}}\,\Phi\!\left(\frac{m}{s(t)}\right),
\qquad
\Phi(\xi)=M_1\,e^{-\xi},
\qquad
\int_0^\infty \xi\,\Phi(\xi)\,d\xi=M_1.
\label{eq:constant_kernel_scaling_function}
\end{equation}

This provides a genuine check on the similarity equation, and it fixes the normalization
question raised in Sec.~\ref{sec:phi_equation}. Substituting $\Phi=M_1e^{-\xi}$ and
$\kappa=1$ into Eq.~\eqref{eq:phi_similarity_equation}, the left side is
$w M_1(2-\xi)e^{-\xi}$ and the right side is
$M_1^2 e^{-\xi}-\tfrac12M_1^2\xi e^{-\xi}=\tfrac{M_1^2}{2}(2-\xi)e^{-\xi}$, so the equation is
satisfied if and only if
\begin{equation}
w=\frac{M_1}{2},
\end{equation}
which is exactly $\dot s$ from Eq.~\eqref{eq:constant_kernel_s} at $\lambda=0$. The amplitude
of $\Phi$ and the coefficient of the growth law are thus locked together, as anticipated: had
we written $z=1$ in place of $w$ on the left-hand side, Eq.~\eqref{eq:phi_similarity_equation}
would have forced $\Phi=2e^{-\xi}$ and hence $M_1=2$ rather than leaving the normalization free.

%=====================================================================
\subsection{Dynamical regimes}\label{sec:regimes}
%=====================================================================

The organizing quantity for the conservative homogeneous problem is the degree
$\lambda=-(\alpha+2\beta)$ together with the accumulated coagulation time $T(t)$. Homogeneity
sets the characteristic-size scaling when a mass-conserving self-similar regime exists, but it
does not determine endpoint exponents, the detailed tail shape, or whether a marginal kernel
gels.

\paragraph{Nongelling kernels, $\lambda<1$.}
The characteristic scale grows as $s\propto T^{1/(1-\lambda)}$. For the one-parameter family
$K\propto(m_1+m_2)^{-\alpha}$ one has $\lambda=-\alpha$, so the marginal boundary $\lambda=1$
lies at $\alpha=-1$, and all kernels simulated below, which have $\alpha,\beta\ge0$, are
comfortably inside the suppressive nongelling region $\lambda\le0$. Their growth exponent for
$\delta=0$ is $z=1/(1+\alpha+2\beta)$.

\paragraph{Marginal degree, $\lambda=1$.}
The algebraic relation $s\propto T^{1/(1-\lambda)}$ ceases to apply and the behavior depends on
the full kernel. The additive kernel $K=m_1+m_2$ is a nongelling solvable example whose
characteristic size grows exponentially in $T$, with an exponential large-mass cutoff and a
nontrivial algebraic prefactor; no universal logarithmic spectrum should be assigned to every
$\lambda=1$ kernel.

\paragraph{Runaway and gelling kernels, $\lambda>1$.}
Homogeneous kernels of degree greater than unity are candidates for gelation or runaway growth,
in which a macroscopic object absorbs a finite fraction of the mass and the pre-gel
mass-conserving description eventually fails~\cite{Leyvraz:2003,Escobedo:2002}. The precise
gelation criterion also depends on the kernel's behavior for very unequal masses, so $\lambda>1$
is a strong diagnostic rather than a theorem. When the kernel contains the separable factor
$t^{-\delta}$, gelation is controlled by whether the cumulative clock $T(t)$ reaches the gel
time of the autonomous problem: $\delta>1$ gives a finite total coagulation time and can prevent
a formally gelling kernel from ever gelling, whereas $\delta<0$---as for L3---accelerates the
approach. Table~\ref{tab:scaling_solutions} places E2, L2 and L3 in this class. The kernels
simulated in Sec.~\ref{sec:numerical} do not probe it, and we make no claim of numerical
validation for the runaway regime.

%=====================================================================
\section{Numerical Results}\label{sec:numerical}
%=====================================================================

\subsection{Method}
\label{sec:mc_method}

We solve the discrete coagulation dynamics directly by kinetic Monte Carlo rather than by
solving Eq.~\eqref{eq:phi_similarity_equation}. Each realization begins with $N_0=1500$
equal-mass seeds of mass $m_0$ and evolves by repeated pairwise merging, with the pair $(i,j)$
drawn with probability proportional to
\begin{equation}
K(i,j)=(m_i+m_j)^{-\alpha}(m_im_j)^{-\beta},\qquad \delta=0,
\label{eq:mc_kernel}
\end{equation}
and results averaged over $10^4$ realizations. The rescaled profile is accumulated in
logarithmic mass bins and fitted to Eq.~\eqref{eq:phi_interpolating} over the range in which the
binned profile is populated.

In the elastic case ($L=1$) a merger produces a remnant of mass $m_i+m_j$; in the inelastic case
($L=0.95$) the remnant has mass $0.95(m_i+m_j)$, the missing $5\%$ being radiated in
gravitational waves. In the figures the radiated fraction is denoted $\mu\equiv1-L=0.05$ in the panel titles of the
mass-loss runs; this $\mu$ is the radiated fraction and is unrelated to the reduced mass
$\mu=m_1m_2/M$ of Eq.~\eqref{eq:psi_def}, which appears only in Sec.~\ref{sec:pbh}.

We sample thirteen exponent pairs: the origin and the two coordinate axes in increments of
$0.2$ up to $1.2$, spanning conservative homogeneities $-2.4\le\lambda\le0$.

\paragraph{Stopping criterion.}
All runs are stopped at a fixed surviving number, $N=N_0/5.435\simeq276$, i.e.\ after $1224$
mergers, so that every kernel is compared at the same stage of coarsening rather than at the
same physical time. This is why the elastic runs all report the same characteristic mass
$s=M_1/N=5.435\,m_0$ while the inelastic runs report smaller values, $s\simeq4.59$--$4.73\,m_0$,
reflecting the mass radiated along the way. One consequence should be stated plainly: because
each kernel is represented by a single snapshot at fixed $N$, these runs constrain the profile
\emph{shape} and carry no information about the growth exponent $z$. Measuring $z$ requires a
time sequence, which we have not performed.

\paragraph{Profile convention.}
Following the figures we plot $\widehat\Phi(\xi;t)=c(m,t)s(t)^2$ against $\xi=m/s(t)$ and fit
Eq.~\eqref{eq:phi_interpolating}. For an elastic similarity solution this ordinate approaches
the time-independent $\Phi$ of Sec.~\ref{sec:scaling}. For $L<1$ it is a snapshot convention
rather than an assertion that $cs^2$ is time independent: indeed
$\int\xi\widehat\Phi\,d\xi=M_1(t)$, which decays according to
Eq.~\eqref{eq:gw_mass_moment_loss}. A shape comparison with the declining mass factored out
would use $cs^2/M_1(t)$; this leaves the fitted $p,q,\xi_0$ unchanged at a given snapshot and
rescales only the amplitude.

\subsection{Elastic profiles and the constant-kernel benchmark}
\label{sec:elastic_results}

\begin{figure}[p]\centering
\begin{subfigure}[t]{0.49\textwidth}\centering\includegraphics[width=\linewidth]{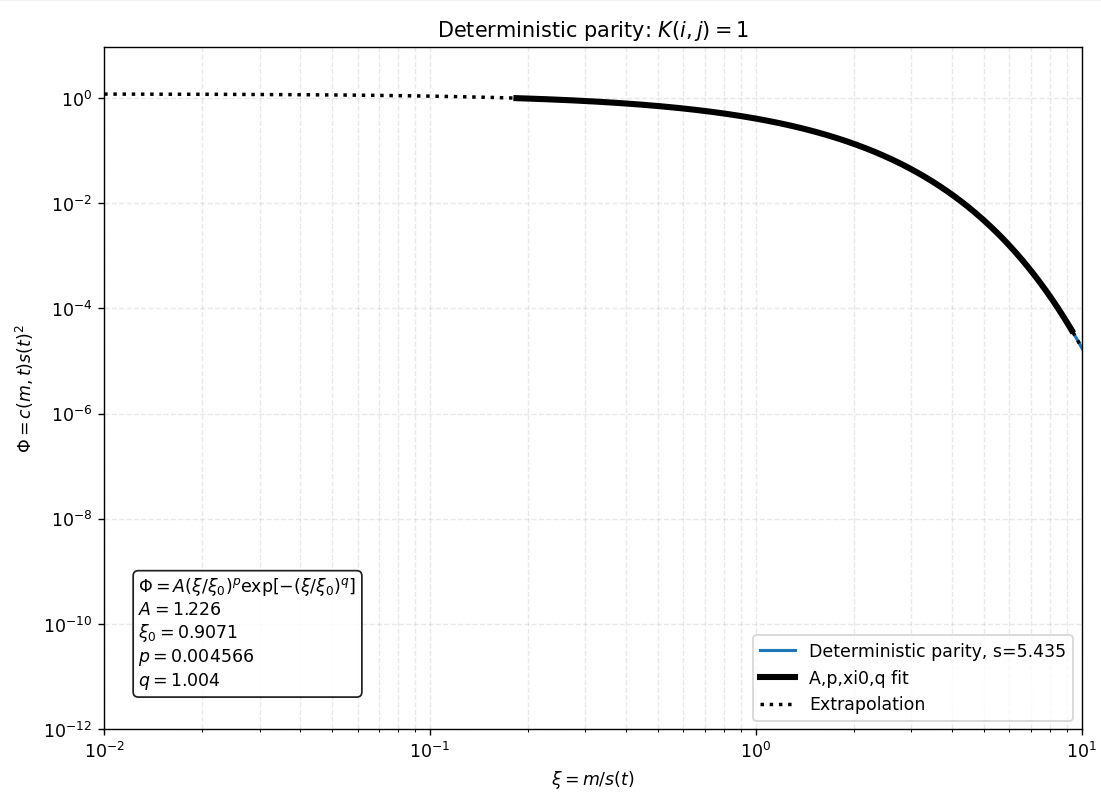}\caption{Elastic, $(0,0)$}\end{subfigure}\hfill
\begin{subfigure}[t]{0.49\textwidth}\centering\includegraphics[width=\linewidth]{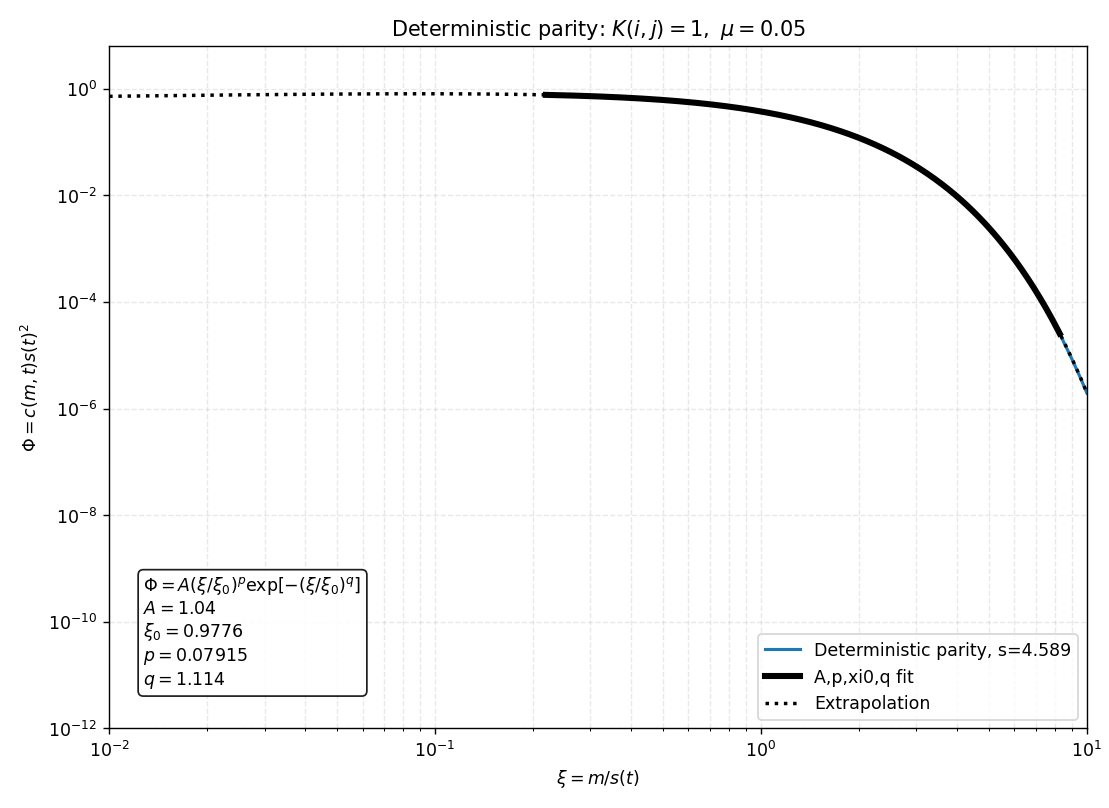}\caption{Mass loss, $(0,0)$}\end{subfigure}\par\medskip
\begin{subfigure}[t]{0.49\textwidth}\centering\includegraphics[width=\linewidth]{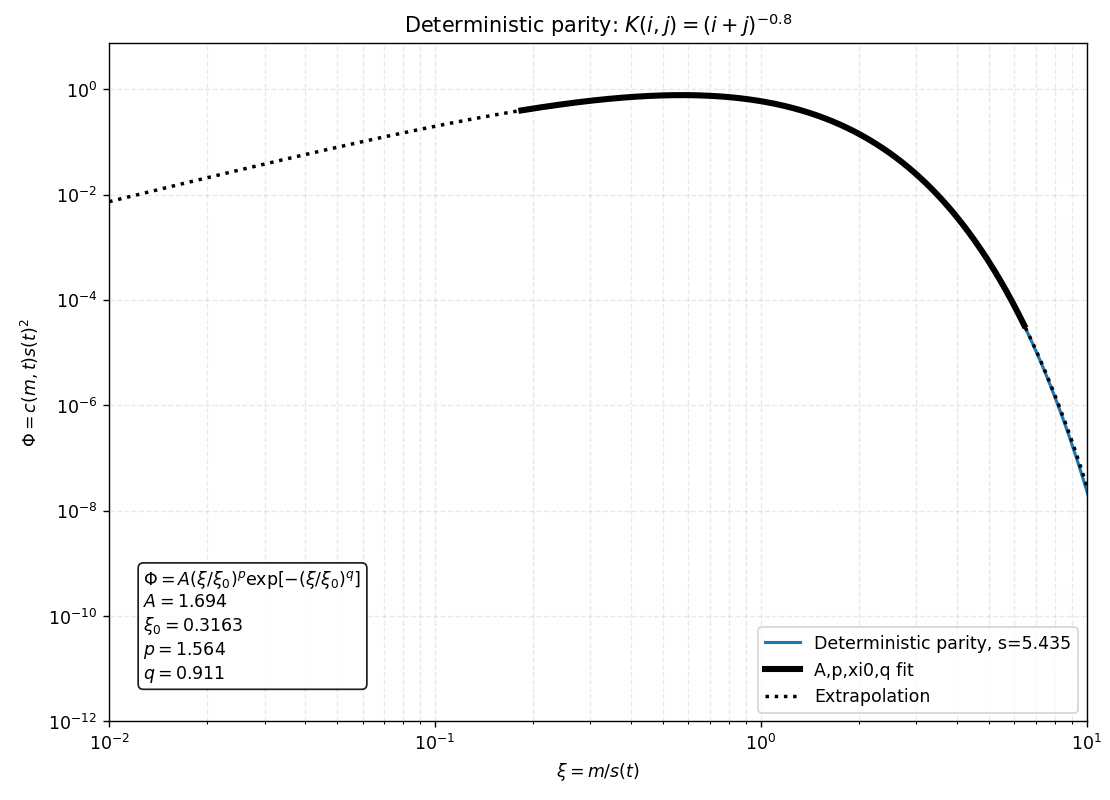}\caption{Elastic, $(0.8,0)$}\end{subfigure}\hfill
\begin{subfigure}[t]{0.49\textwidth}\centering\includegraphics[width=\linewidth]{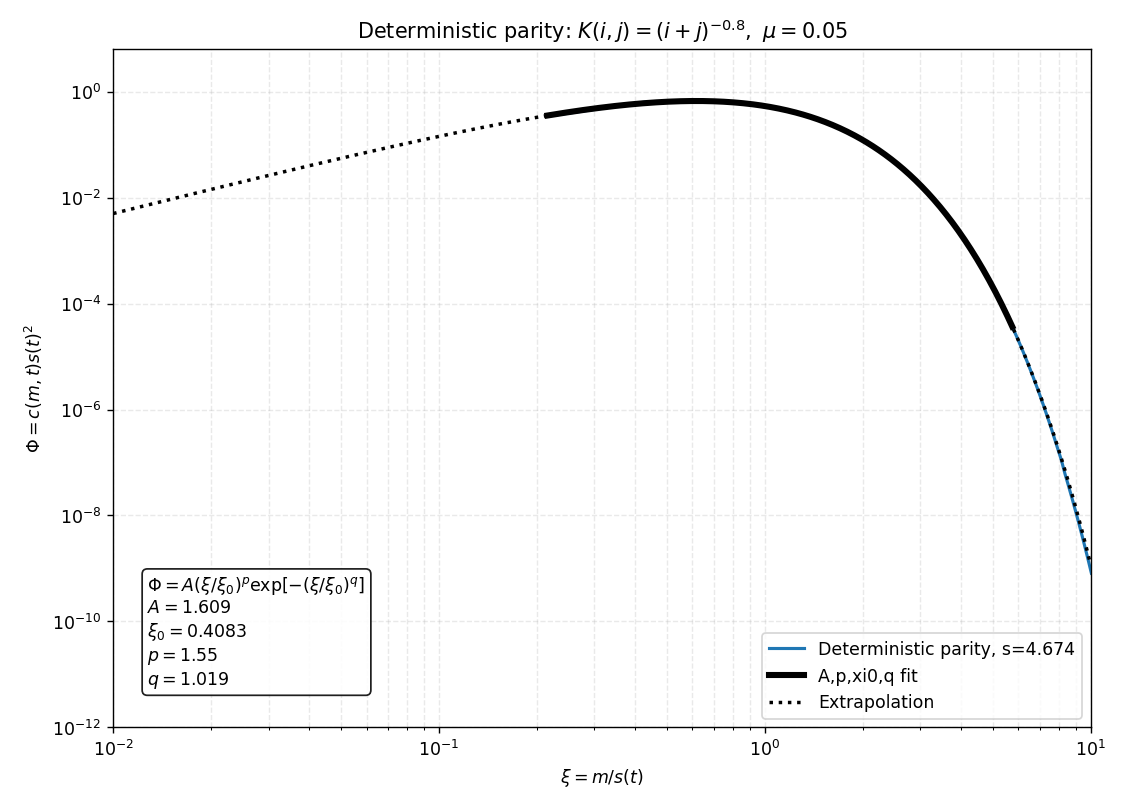}\caption{Mass loss, $(0.8,0)$}\end{subfigure}\par\medskip
\begin{subfigure}[t]{0.49\textwidth}\centering\includegraphics[width=\linewidth]{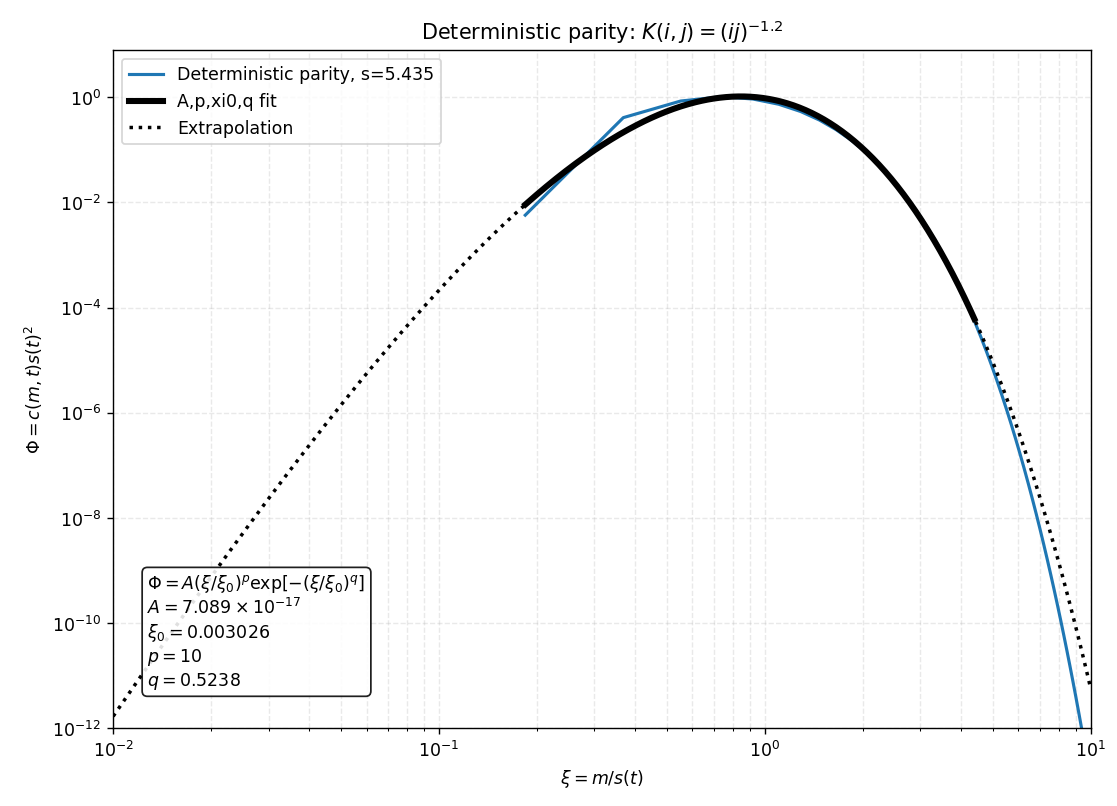}\caption{Elastic, $(0,1.2)$}\end{subfigure}\hfill
\begin{subfigure}[t]{0.49\textwidth}\centering\includegraphics[width=\linewidth]{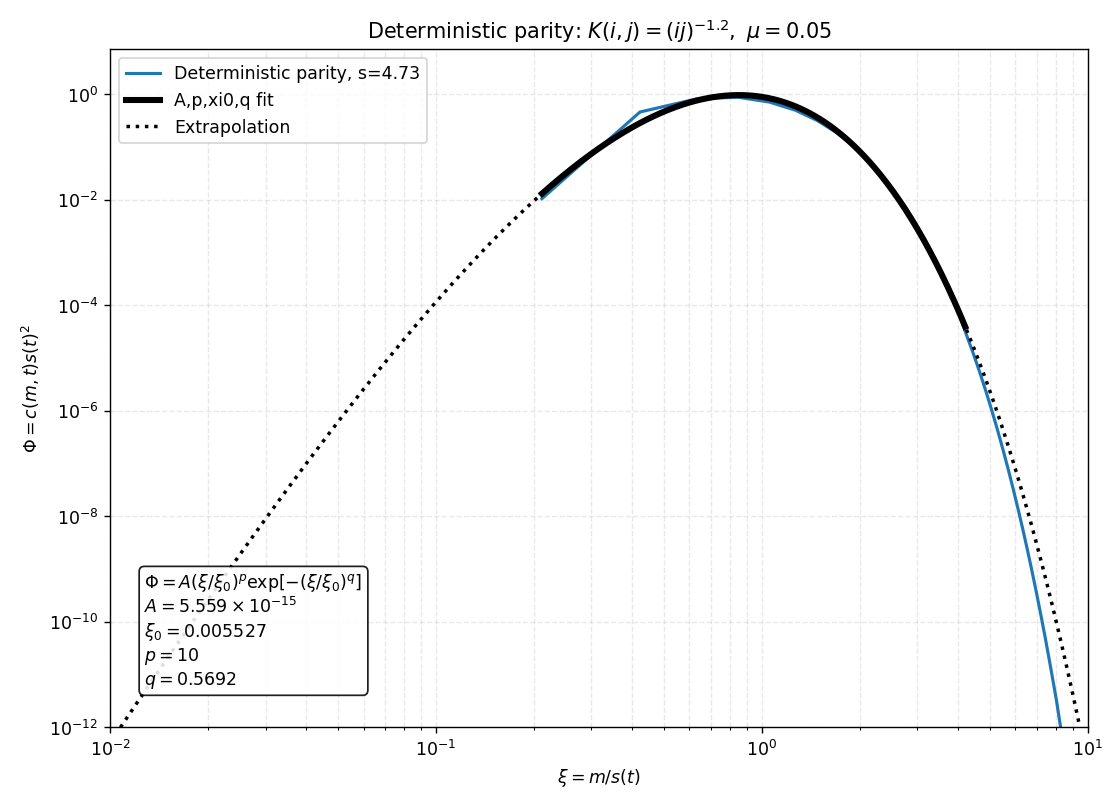}\caption{Mass loss, $(0,1.2)$}\end{subfigure}
\caption{Rescaled profiles $\widehat\Phi=c\,s^2$ for the constant kernel (top), a
total-mass-suppressive kernel (middle), and a product-mass-suppressive kernel (bottom), without
(left) and with (right) a $5\%$ radiative loss at each merger. Blue curves are the Monte Carlo
profiles, solid black curves the fits of Eq.~\eqref{eq:phi_interpolating} over the fitted
interval, and dotted black curves their extrapolation beyond it. For $L<1$ the ordinate does not
remove the declining total mass.}
\label{fig:profiles}\end{figure}

Table~\ref{tab:MC_table} gives the four-parameter fits and Fig.~\ref{fig:profiles}
shows representative profiles. At $(\alpha,\beta)=(0,0)$ the elastic fit gives $p=0.0046$ and
$q=1.004$, close to the exact exponential shape $p=0$, $q=1$ derived in
Sec.~\ref{subsec:constant-kernel-exact}. This is a useful consistency check on the Monte Carlo,
though at a single snapshot and finite population it is not a test of convergence.

\begin{figure}[htbp]\centering
\begin{subfigure}[t]{0.49\textwidth}\centering\includegraphics[width=\linewidth]{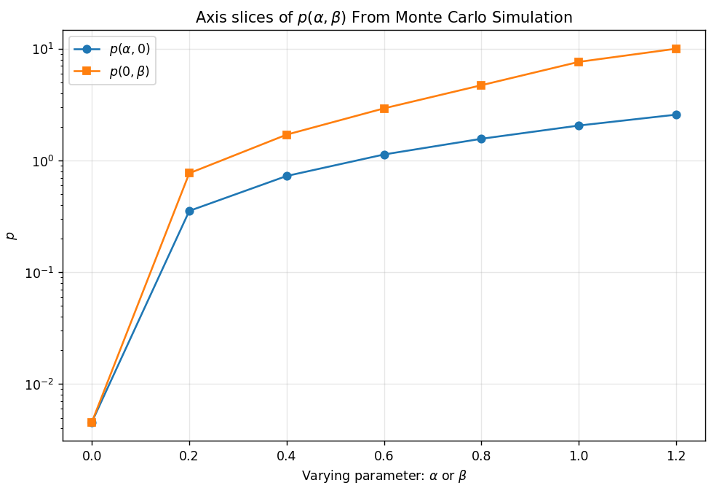}\caption{Elastic $p$}\end{subfigure}\hfill
\begin{subfigure}[t]{0.49\textwidth}\centering\includegraphics[width=\linewidth]{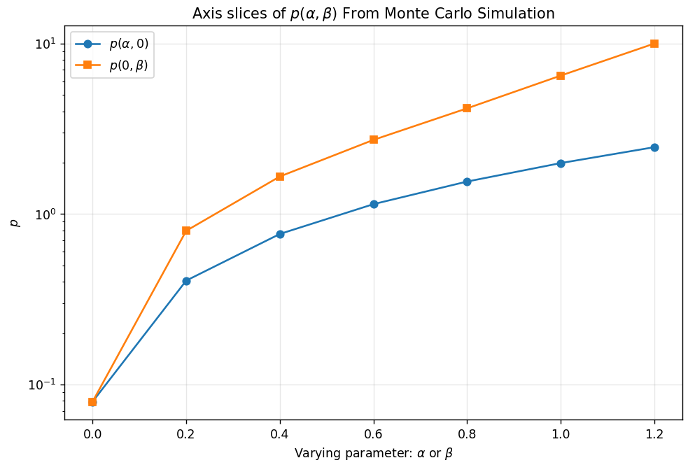}\caption{Mass-loss $p$}\end{subfigure}\par\medskip
\begin{subfigure}[t]{0.49\textwidth}\centering\includegraphics[width=\linewidth]{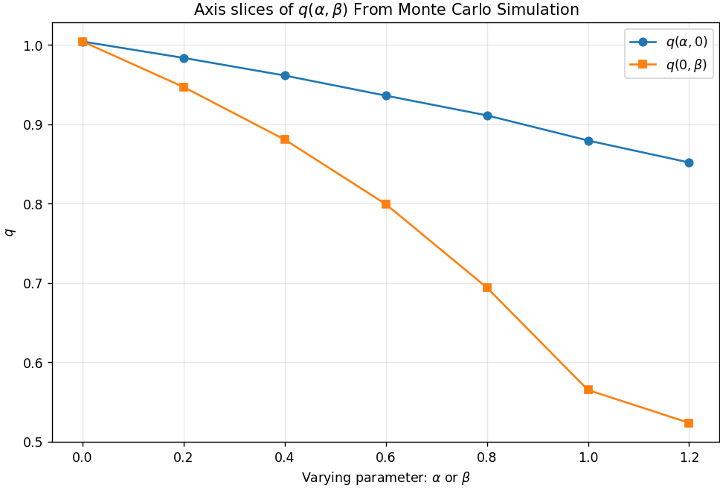}\caption{Elastic $q$}\end{subfigure}\hfill
\begin{subfigure}[t]{0.49\textwidth}\centering\includegraphics[width=\linewidth]{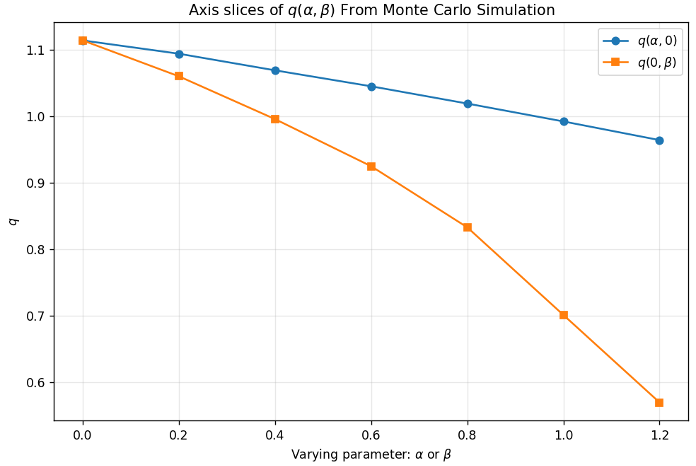}\caption{Mass-loss $q$}\end{subfigure}
\caption{Axis slices of the fitted profile parameters. Blue circles vary $\alpha$ at $\beta=0$;
orange squares vary $\beta$ at $\alpha=0$. The $p$ panels use logarithmic vertical axes. Lines
connect the sampled points and are not fitted response laws.}
\label{fig:axes}\end{figure}

Along the $\alpha$ axis, $p$ rises to $2.57$ while $q$ falls to $0.85$. Along the $\beta$ axis
the response is much stronger: at $\beta=1.2$, $q=0.52$ and the fitted $p$ reaches the optimizer
bound. The scale parameter falls from $\xi_0=0.91$ at the origin to $0.18$ at $\alpha=1.2$ and
to $0.0030$ at $\beta=1.2$. Figure~\ref{fig:axes} displays the contrasting axis responses.

These differences are not captured by homogeneity alone. The kernels $(\alpha,\beta)=(0.8,0)$
and $(0,0.4)$ both have $\lambda=-0.8$, yet their elastic fits give $(p,q)=(1.56,0.911)$ and
$(1.70,0.881)$ respectively. This is the kernel-shape dependence anticipated in
Sec.~\ref{sec:phi_tail}: total-mass suppression and product-mass suppression are not
interchangeable even at fixed $\lambda$. Absent fit covariances we do not attach a significance
to the size of the difference.

\begin{table}[!htbp]\centering
\begin{tabular}{rr rr rr rr rr}\toprule
& & \multicolumn{2}{c}{$A$} & \multicolumn{2}{c}{$\xi_0$}
& \multicolumn{2}{c}{$p$} & \multicolumn{2}{c}{$q$}\\
\cmidrule(lr){3-4}\cmidrule(lr){5-6}\cmidrule(lr){7-8}\cmidrule(lr){9-10}
$\alpha$ & $\beta$ & $L{=}1$ & $L{=}0.95$ & $L{=}1$ & $L{=}0.95$
& $L{=}1$ & $L{=}0.95$ & $L{=}1$ & $L{=}0.95$\\\midrule
$0$ & $0$ & $1.23$ & $1.04$ & $0.907$ & $0.978$ & $0.0046$ & $0.079$ & $1.004$ & $1.114$\\
$0.2$ & $0$ & $1.69$ & $1.39$ & $0.685$ & $0.779$ & $0.354$ & $0.406$ & $0.983$ & $1.094$\\
$0.4$ & $0$ & $2.00$ & $1.67$ & $0.527$ & $0.623$ & $0.727$ & $0.763$ & $0.961$ & $1.069$\\
$0.6$ & $0$ & $2.01$ & $1.75$ & $0.407$ & $0.504$ & $1.13$ & $1.14$ & $0.936$ & $1.045$\\
$0.8$ & $0$ & $1.69$ & $1.61$ & $0.316$ & $0.408$ & $1.56$ & $1.55$ & $0.911$ & $1.019$\\
$1$ & $0$ & $1.13$ & $1.27$ & $0.240$ & $0.331$ & $2.05$ & $1.99$ & $0.879$ & $0.992$\\
$1.2$ & $0$ & $0.599$ & $0.836$ & $0.184$ & $0.267$ & $2.57$ & $2.47$ & $0.852$ & $0.964$\\[2pt]
$0$ & $0.2$ & $2.02$ & $1.68$ & $0.501$ & $0.605$ & $0.769$ & $0.797$ & $0.947$ & $1.060$\\
$0$ & $0.4$ & $1.49$ & $1.50$ & $0.276$ & $0.375$ & $1.70$ & $1.66$ & $0.881$ & $0.996$\\
$0$ & $0.6$ & $0.277$ & $0.570$ & $0.135$ & $0.224$ & $2.93$ & $2.72$ & $0.799$ & $0.925$\\
$0$ & $0.8$ & $1.8{\times}10^{-3}$ & $3.7{\times}10^{-2}$ & $0.0476$ & $0.113$ & $4.71$ & $4.17$ & $0.694$ & $0.833$\\
$0$ & $1$ & $4.1{\times}10^{-10}$ & $1.0{\times}10^{-5}$ & $0.0080$ & $0.0343$ & $7.63$ & $6.49$ & $0.565$ & $0.701$\\
$0$ & $1.2$ & $7.1{\times}10^{-17}$ & $5.6{\times}10^{-15}$ & $0.0030$ & $0.0055$ & $10^{*}$ & $10^{*}$ & $0.524$ & $0.569$\\
\bottomrule\end{tabular}
\caption{Profile fits to Eq.~\eqref{eq:phi_interpolating} for the elastic ($L=1$) and
gravitational-wave mass-loss ($L=0.95$) runs. Asterisks mark fits in which $p$ reached the
optimizer bound; at those points $A$ and $\xi_0$ are strongly degenerate and only the
combination $A\xi_0^{-p}$ is determined over the fitted interval (see
Sec.~\ref{sec:scope}). Fit covariances are not reported.}
\label{tab:MC_table}\end{table}

\subsection{Effect of gravitational-wave mass loss}
\label{sec:massloss_results}

The mass-loss profiles retain the same qualitative axis trends, but their fitted parameters do
not coincide with the elastic values, and the exact conservative exponential is a benchmark for
the $L=1$ calculation only: at the constant kernel the inelastic fit gives $(p,q)=(0.079,1.114)$.
Along the $\alpha$ axis $q$ remains above unity through $\alpha=0.8$ and falls to $0.964$ at
$\alpha=1.2$; along the $\beta$ axis it falls to $0.569$.

The most systematic effect is that $q_{\rm I}>q_{\rm E}$ at every sampled pair, by about $0.11$
along the $\alpha$ axis and by between $0.045$ and $0.139$ along the $\beta$ axis. The fitted
$\xi_0$ is likewise larger in the mass-loss case at every pair. The response of $p$ is not
uniform in sign: mass loss raises it near the origin and at small exponents but lowers it beyond
a crossover that comes later along $\alpha$ (between $\alpha=0.6$ and $0.8$) than along $\beta$
(by $\beta=0.4$), before both $\beta=1.2$ fits reach the optimizer bound.

Two features of this comparison deserve emphasis, because they are the clearest dynamical
signature of the non-conservative nature of the problem. First, the shifts are one-directional:
$q$ and $\xi_0$ increase at all thirteen kernels, elastic to inelastic, without exception. A
random or purely statistical difference between two independent sets of runs would not do this.
Second, the size of the shift in $q$ is nearly independent of $\alpha$ (about $0.11$ across the
whole axis) but varies by a factor of three along $\beta$, peaking at $\beta=0.8$ and collapsing
to $0.045$ at $\beta=1.2$ where both fits reach the optimizer bound. Mass loss therefore
interacts with product-mass suppression in a way it does not with total-mass suppression. Because
the retained fraction is held fixed at $L=0.95$ for every merger in this study, the effect cannot
be attributed to any mass-ratio dependence of the radiated energy; it must instead reflect how
the $(m_1m_2)^{-\beta}$ suppression reshapes which pairs merge and in what order, i.e.\ the
composition of the merger tree. We report this as an empirical feature of the simulated kernels
rather than derive it from first principles.

Underlying both is the moment equation~\eqref{eq:gw_mass_moment_loss}: the population is losing
black-hole mass continuously, at a rate proportional to the same kernel that drives the
coarsening. The conservative relation $\theta=2z$ therefore does not survive, and the amplitude
and characteristic-scale exponents must be measured independently rather than inherited from the
$L=1$ theory. Profile shapes remain close enough that a similarity description is useful over
restricted ranges, but it is a two-exponent description rather than a one-exponent one.

A larger $q$ by itself does not establish stronger tail suppression at fixed physical mass,
because the amplitude, scale and algebraic factor all change together. From
Eq.~\eqref{eq:phi_interpolating},
\begin{equation}
\frac{d\ln\widehat\Phi}{d\ln\xi}=p-q(\xi/\xi_0)^q,
\qquad
\xi_{\rm peak}=\xi_0(p/q)^{1/q}\quad(p>0),
\label{eq:mc_local_slope}
\end{equation}
so an increasing $p$ and decreasing $q$ can offset one another. Comparing populations requires
the full profile together with the measured $s(t)$ and its normalization, and---for $L<1$---the
mass-moment equation~\eqref{eq:gw_mass_moment_loss}.

\subsection{Parameter surfaces and the role of homogeneity}
\label{sec:surfaces}

\begin{figure}[htbp]\centering
\begin{subfigure}[t]{0.49\textwidth}\centering\includegraphics[width=\linewidth]{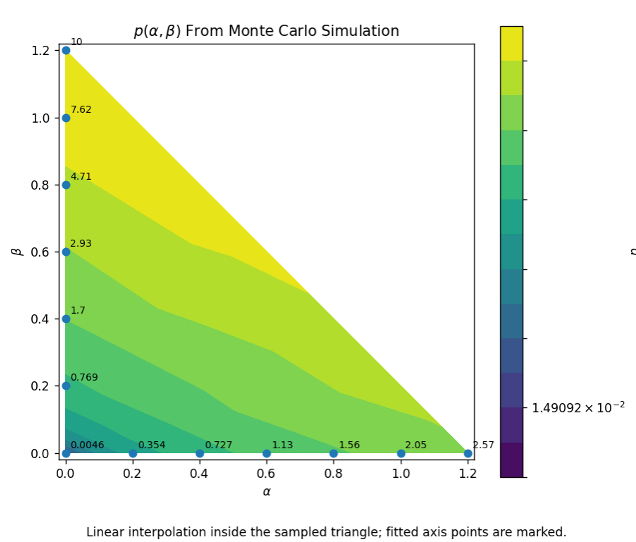}\caption{$p$, elastic ($L=1$)}\end{subfigure}\hfill
\begin{subfigure}[t]{0.49\textwidth}\centering\includegraphics[width=\linewidth]{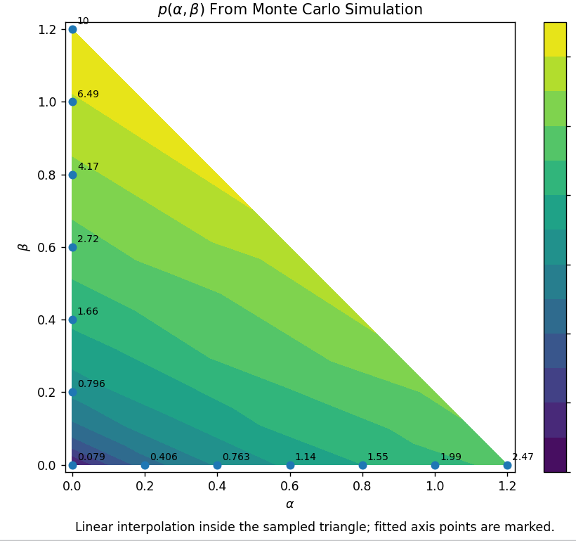}\caption{$p$, mass loss ($L=0.95$)}\end{subfigure}\par\medskip
\begin{subfigure}[t]{0.49\textwidth}\centering\includegraphics[width=\linewidth]{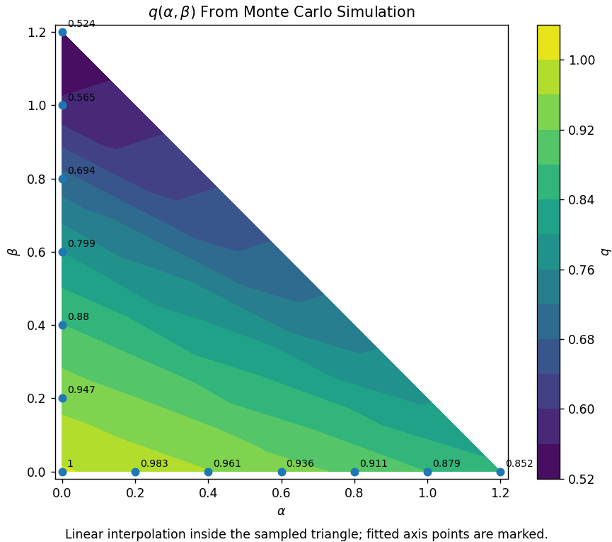}\caption{$q$, elastic ($L=1$)}\end{subfigure}\hfill
\begin{subfigure}[t]{0.49\textwidth}\centering\includegraphics[width=\linewidth]{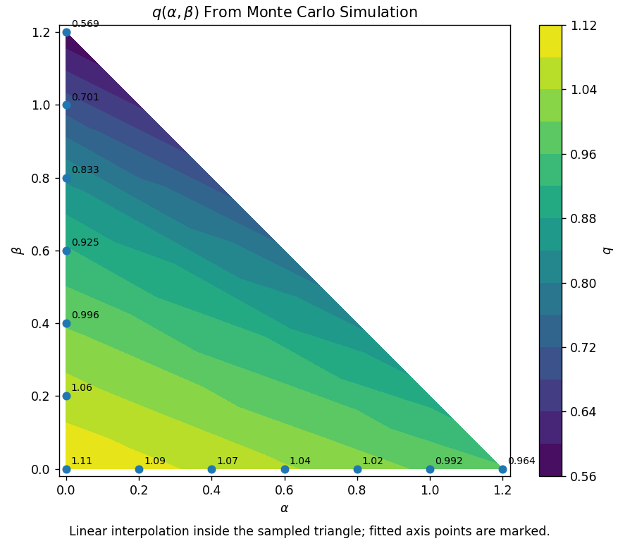}\caption{$q$, mass loss ($L=0.95$)}\end{subfigure}
\caption{Fitted profile exponents over the sampled $(\alpha,\beta)$ region, without (left) and
with (right) a $5\%$ radiative loss at each merger. Markers indicate the thirteen kernels
actually solved; the colored interior is a linear interpolation within their triangular convex
hull. Contours of constant $p$ and $q$ are steeper than the iso-homogeneity lines
$\alpha+2\beta=\text{const}$, indicating that the profile shape is not a function of $\lambda$
alone.}
\label{fig:contours}\end{figure}

Figure~\ref{fig:contours} presents the fitted exponents as surfaces over the sampled
$(\alpha,\beta)$ region, for both the elastic and the mass-loss runs. Both $p$ and $q$ vary
monotonically and in opposite senses: $p$ increases and $q$ decreases as either suppressive
exponent is raised, with the largest values of $p$ and the smallest values of $q$ in the upper
part of the region, where product-mass suppression is strongest. The two families of surfaces
are qualitatively alike, and the effect of radiative losses appears as a nearly rigid offset of
the $q$ surface toward larger values, consistent with the pointwise comparison of
Sec.~\ref{sec:massloss_results}.

The surfaces are useful for one specific question, which is whether the homogeneity degree
organizes the profile shape. If $p$ and $q$ were functions of $\lambda$ alone, their contours
would coincide with the iso-homogeneity lines $\alpha+2\beta=\text{const}$, which are straight
lines of slope $-1/2$ in the $(\alpha,\beta)$ plane. They do not. Along the $\beta$ axis a unit
step changes $\lambda$ twice as fast as along the $\alpha$ axis, so a $\lambda$-controlled
response would show the $\beta$ axis rising exactly twice as steeply; instead $p$ rises from
$0.005$ to $10$ along $\beta$ against $0.005$ to $2.57$ along $\alpha$, roughly a factor of four.
The contours are correspondingly steeper than the iso-$\lambda$ lines at every sampled axis
point (the only points these surfaces are anchored to; see below). This is the same conclusion reached pointwise from the equal-homogeneity pair
$(0.8,0)$ and $(0,0.4)$ in Sec.~\ref{sec:elastic_results}, and it is the numerical counterpart
of the analytic statement in Sec.~\ref{sec:phi_tail} that no relation $q=q(\lambda)$ holds for
this kernel class.

We stress that this comparison rests on the sampled axis points and not on the colored interiors,
which are linear interpolations within the triangular convex hull rather than additional
numerical runs. Detailed bends or curvature visible inside the triangle are artifacts of that
interpolation, and no mixed-parameter interaction should be read from them; establishing one
would require solving interior kernels directly.

\subsection{Scope and limitations}
\label{sec:scope}

Three limitations bound the numerical conclusions. First, only coordinate-axis kernels were
sampled, so the response to mixed exponents is untested; the equal-$\lambda$ comparison above
shows that such a test would be worthwhile. Second, each kernel is represented by a single
snapshot at fixed surviving number, so no growth exponent is measured and no self-similar
collapse in time is demonstrated; the runs also do not probe the marginal $\lambda=1$ or gelling
$\lambda>1$ regimes into which Table~\ref{tab:scaling_solutions} places three of the four PBH
channels. Third, the four-parameter fits are not everywhere well constrained. The fitted
interval spans roughly $0.18\lesssim\xi\lesssim10$, so at the strongly suppressive $\beta$ values
where $\xi_0\ll0.18$ the fit never samples $\xi\lesssim\xi_0$; there $A(\xi/\xi_0)^p$ reduces to
a single power law and only the combination $A\xi_0^{-p}$ is determined, which is why $p$ runs
to its bound at $\beta=1.2$ in both the elastic and mass-loss fits. With $N=276$ surviving
objects at the snapshot, the tail near $\xi\sim10$ also rests on few objects. Realization-level
scatter and fit covariances would be needed to attach uncertainties to the trends reported
above, and the dotted extrapolations in Fig.~\ref{fig:profiles} should not be read as measuring
the $\xi\to0$ asymptote.

%=====================================================================
\section{Discussion and Conclusions}\label{sec:conclusions}
%=====================================================================

We have formulated hierarchical black-hole mergers as a Smoluchowski coagulation problem with a
separable homogeneous kernel. For the conservative equation the natural variable is the
cumulative coagulation clock $T(t)=\int^tt'^{-\delta}dt'$, in which the characteristic mass obeys
$s\propto T^{1/(1-\lambda)}$ for degree $\lambda<1$, reducing for $\delta<1$ to $s\propto t^z$
with $z=(1-\delta)/(1+\alpha+2\beta)$ and $c(m,t)=s^{-2}\Phi(m/s)$. The commonly introduced
prefactor $m^{-\tau}$ is not an independent observable exponent, since it can be absorbed into
$\Phi$; we therefore use the $\tau=0$ representation while allowing the profile itself to carry
nontrivial endpoint behavior. The constant kernel provides an exact benchmark: for monodisperse
initial data the discrete spectrum is geometric and approaches $\Phi(\xi)=M_1e^{-\xi}$ with
$s(t)=M_1t/2$, which closes the similarity equation and fixes the relation between the profile
normalization and the growth-law coefficient.

The central bookkeeping result concerns the translation of published PBH merger rates into
coagulation kernels. Rates quoted as
$dR/d\ln m_1d\ln m_2 = C\,M^{-a'}\eta^{-b'}t^{-\delta}\psi(m_1)\psi(m_2)$ are rate
\emph{densities}; extracting the kernel requires dividing by $c(m_1)c(m_2)$, which introduces a
factor $m_1m_2=\eta M^2$ and shifts the exponents to $a=a'-2$, $b=b'-1$, hence
$\lambda=2-a'$. The combination $\alpha=a-2b$ is invariant under this shift, but $\beta$ and
$\lambda$ are not. The consequence is a reclassification of the standard channels: only the
early three-body pathway maps into the nongelling, mass-conserving similarity regime, with
$z\simeq0.0305$ ($\gamma=1$) and $0.0716$
($\gamma=2$), while E2, L2 and L3 all have $\lambda>1$. For L2 this is confirmed independently
by computing $\langle\sigma v\rangle$ from the gravitational-wave capture cross-section, which
gives $\lambda=2$ directly; for E2 it is confirmed in the monochromatic limit, where
$\lambda=42/37$. That the early two-body kernel is superlinear is the coagulation-language
statement of a known feature of PBH pairing, namely that the early-binary rate is dominated by
the heaviest objects present. Whether these formal runaway diagnoses correspond to physical
gelation depends on the kernel at very unequal masses, on the suppression factors $S_E$ and
$S_L$, and on whether the coagulation clock reaches the gel time---none of which is settled here.

Numerically, thirteen kernels were solved by Monte Carlo with and without a $5\%$ radiative loss
at each merger. The elastic constant-kernel fit reproduces the exact exponential shape.
Increasing either suppressive exponent raises $p$ and lowers $q$, with a markedly larger response
along the $\beta$ axis, and kernels of equal homogeneity but different weighting give
different fitted shapes at the level of the point estimates, consistent with $\lambda$ not
determining the profile on its own (we do not report fit uncertainties; see
Sec.~\ref{sec:scope}). Mass loss
raises $q$ at every sampled kernel while also changing the scale and prefactor, so tail
abundances must be compared using the full distribution rather than $q$ alone. Because
gravitational-wave emission makes the process non-conservative in black-hole mass, with $M_1$
decaying according to Eq.~\eqref{eq:gw_mass_moment_loss}, the conservative relation $\theta=2z$
does not carry over; profile-shape similarity may remain useful over restricted ranges, but the
amplitude and characteristic scale must be determined from the non-conservative evolution itself.

Several extensions follow directly. Solving Eq.~\eqref{eq:phi_similarity_equation} for $\Phi$
by fixed-point or collocation methods would provide an independent check on the Monte Carlo
profiles. On the simulation side, the priorities are to sample mixed $(\alpha,\beta)$ kernels in
the nongelling quadrant, to take multiple snapshots so that a genuine self-similar collapse and
a measured growth exponent can be exhibited, and to simulate representative $\lambda>1$ kernels,
including the converted E2 and L2 exponents, with explicit monitoring of moment loss and
finite-size effects. The last of these is now the most consequential: with three of the four PBH
channels placed in the runaway class, the gelling side of the phase diagram is no longer a
peripheral case but the regime in which most of the astrophysical interest lies.

\section*{Data Availability}
The Monte Carlo code and numerical outputs underlying Sec.~\ref{sec:mc_method}--\ref{sec:scope}
are available from the  authors upon reasonable request.

\section*{Acknowledgments}
This work is partly supported by the U.S. Department of Energy
grant number de-sc0010107 (SP).

\clearpage
% Bibliography

%% [A] Recommended: using JHEP.bst file
\bibliographystyle{JHEP}

\bibliography{biblio}
\end{document}